\documentclass[varenna]{cimento}

\setcopyright{Victor V. Albert}
\usepackage{graphicx}  
\graphicspath{{./figures/}}

\usepackage{nicefrac}
\usepackage{amsmath, amssymb, graphicx}
\usepackage[draft]{changes}
\usepackage{dsfont}

\usepackage{cite}

\usepackage{mnsymbol}
\usepackage{booktabs}
\providecommand{\tabularnewline}{\\}

\newcommand{\N}{{\cal{N}}}
\newcommand{\E}{{\cal{E}}}
\newcommand{\D}{{\cal{D}}}

\newcommand{\Z}{{\mathbb{Z}}}
\newcommand{\R}{{\mathbb{R}}}
\newcommand{\EE}{\hat{E}}

\newcommand{\ket}{\rangle}
\newcommand{\bra}{\langle}
\newcommand{\dg}{{\dagger}}
\newcommand{\der}{{\text{d}}}
\newcommand{\xx}{{\hat{x}}}
\newcommand{\pp}{{\hat{p}}}
\newcommand{\al}{{\hat{a}}}
\newcommand{\nh}{{\hat{n}}}
\newcommand{\cnot}{{\textsc{cnot}}}

\global\long\def\df#1{\textbf{#1}}
\global\long\def\zoo#1{
	\mbox{\pdfstartlink attr{/Border[0 0 0]} user{/Subtype /Link /A << /S /URI /URI (https://errorcorrectionzoo.org/c/#1) >>}{\includegraphics[height=10pt,keepaspectratio]{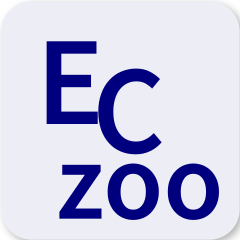}}\pdfendlink}\!
}

\DeclareMathOperator*{\argmin}{{\text{argmin}}}

\title{Bosonic coding: introduction and use cases}

\author{Victor~V.~Albert}

\institute{Joint Center for Quantum Information and Computer Science\\
NIST and University of Maryland\\
4254 Stadium Dr, College Park, MD 20740, USA}

\begin{document}

\maketitle

\begin{abstract}
Bosonic or continuous-variable coding
is a field concerned with robust quantum information processing and communication with electromagnetic signals or mechanical modes.
I review bosonic quantum memories,
characterizing them as either bosonic stabilizer or bosonic Fock-state codes.
I then enumerate various applications of bosonic encodings, four of which circumvent no-go theorems due to the intrinsic infinite-dimensionality of bosonic systems.
\end{abstract}

\section{Introduction \& outline}
\label{sec:qec}

Noise is pervasive in both classical and quantum devices, and the goal of error correction is to preserve messages sent through a noisy transmission channel.
This is usually done by encoding the messages into an error-correcting code~\cite{Shannon1948}\zoo{ecc}.

A robust encoding of (classical) information into the frequency, amplitude, or phase of one or more electromagnetic signals is called a coded modulation scheme~\cite{Anderson2002,Lapidoth2017}\zoo{analog}.
For example, one can encode the two values of a bit into temporal waveforms \(\pm\cos t\) of opposite phase, corresponding to the binary phase-shift keyring (PSK) modulation scheme.
The task of bosonic coding\zoo{oscillators} is to provide robust encodings of \textit{quantum} information into electromagnetic signals (see Table~\ref{tab:comparison}).

\begin{table}[]{}
\begin{tabular}{cc}
\toprule
Classical paradigm & Quantum paradigm\tabularnewline
\midrule
coded modulation scheme & bosonic encoding\tabularnewline
bitstream or baseband & logical information\tabularnewline
spectrum & number of modes\tabularnewline
bandwidth efficiency & logical information per mode\tabularnewline
\midrule
quadrature amplitude modulation & GKP code\tabularnewline
quadrature phase-shift keyring & cat code\tabularnewline
frequency-shift keyring & frequency-bin dual-rail code\tabularnewline
\bottomrule
\end{tabular}

\caption{
	Side-by-side comparison of related terminology describing error-correcting encodings of classical and quantum information into electromagnetic signals.
	The first four terms apply to general encodings, while the last three represent parallels between modulation schemes~\cite{Anderson2002} and some of the bosonic encodings discussed here; two such parallels were first noted in Ref.~\cite{Girvin2021}.
	More details can be found on the corresponding code pages of the error-correction zoo~\mbox{\pdfstartlink attr{/Border[0 0 0]} user{/Subtype /Link /A << /S /URI /URI (https://errorcorrectionzoo.org) >>}{\includegraphics[height=10pt,keepaspectratio]{zoo_icon.png}}\pdfendlink}.
	\label{tab:comparison}}
\end{table}

This brief review of bosonic codes assumes some basic knowledge of qubit-based error correction, in particular qubit stabilizer codes, which I tersely summarize in Sec.~\ref{sec:qubits} before introducing bosonic modes in Sec.~\ref{sec:modes}.
There, I advocate for the use of Schwartz space over Hilbert space for defining bosonic quantum states.

The bosonic codes I review are summarized in Table~\ref{tab:codes}.
They are grouped into two classes --- bosonic stabilizer codes (Sec.~\ref{sec:bosonic-stab}) and bosonic Fock-state codes (Sec.~\ref{sec:fock-state}).
The former generalize the framework of qubit stabilizer codes and can be formulated naturally using what is known as the ``position basis''.
The latter start to stray away from the stabilizer framework and can be formulated naturally using so-called Fock states.
I discuss how codes fare against common noise channels in Sec.~\ref{sec:physical-noise}.

I list various use cases for bosonic codes and, more generally, infinite-dimensional spaces in Sec.~\ref{sec:advantage}.
Bosonic codes can utilize available hardware in a more efficient way than finite-dimensional discrete-variable (DV) codes, can help improve performance of DV codes via concatenation, allow for continuous-parameter families of certain gates that are forbidden in the DV world, provide a way to circumvent magic-state distillation, and can be helpful in simulating bosonic topological phases, among other bosonic systems.
I ponder the future in Sec.~\ref{sec:summary}.

Readers interested in more extensive reviews are welcome to peruse related expositions on qubit-based error correction~\cite{preskillnotes,Nielsen2011,Terhal2015,gottesmanbook}, certain bosonic codes~\cite{Joshi2021,Cai2020,Terhal2020,Noh2021a,Guillaud2022}, and an excellent mix of both~\cite{Girvin2021}.
I also place relevant clickable links to parts of the error-correction zoo, using the icon~\mbox{\pdfstartlink attr{/Border[0 0 0]} user{/Subtype /Link /A << /S /URI /URI (https://errorcorrectionzoo.org) >>}{\includegraphics[height=10pt,keepaspectratio]{zoo_icon.png}}\pdfendlink}, next to codes or topics as they are introduced.
Introductions to bosonic systems can be found in the books~\cite{Cerf2007,serafinibook}.

\begin{table}[]{}
 \centering
\begin{tabular}{ccc}
\toprule
Bosonic code & Encoding & Check operators \tabularnewline
\midrule
analog stabilizer & logical modes & nullifiers\tabularnewline
GKP & logical qudits & modular position \& modular momentum\tabularnewline
GKP-stabilizer & logical modes & modular position \& modular momentum\tabularnewline
\midrule
number-phase & logical qudits & modular number \& modular phase\tabularnewline
cat & logical qudits & modular number \tabularnewline
binomial \& Chebyshev & logical qudits & modular number \tabularnewline
dual-rail & logical qubit & number (error-detecting only) \tabularnewline
CLY & logical qudit & number \tabularnewline
\bottomrule
\end{tabular}
\caption{Some of the bosonic encodings discussed in this brief review, along with their check operators. The first three are bosonic stabilizer codes (see Sec.~\ref{sec:bosonic-stab}), and the last five are bosonic Fock-state codes (see Sec.~\ref{sec:fock-state}). The dimension of the logical space storing the message \(\rho\) for bosonic codes may be finite, corresponding to a logical-qubit or qudit encoding, or infinite, corresponding to an analog or logical-mode encoding.
}
\label{tab:codes}
\end{table}

\section{Conventional quantum error correction}\label{sec:qubits}

In an error-correcting protocol, a message \(\rho\) is encoded by the sender (\textit{a.k.a.}\@ source) using an encoding map \(\E\), passed through a noisy channel \(\N\), and extracted by the receiver (\textit{a.k.a.}\@ sink) using a decoding map \(\D\).
The noisy channel can represent transmission through space
or storage in a stationary memory device.
If error correction succeeds, then the message should remain roughly intact such that noisy transmission is nearly equivalent to the noiseless case,
\begin{equation}\label{eq:qec}
\D\N\E(\rho) \approx \rho\quad\quad\text{(error correction).}
\end{equation}
The same picture holds irrespective of whether the information is classical or quantum, with the difference lying in the structure of the message and the inputs and outputs of the maps \(\E\), \(\N\) and \(\D\).

In a computational setting, one is also interested in performing
operations, or \df{logical gates}, on the stored information.
The subfield of fault tolerance~\cite{Gottesman2009} is concerned with successfully applying logical gates and other operations on the message within a noisy environment.

The smallest non-trivial quantum system is a two-level system, \textit{a.k.a.}\@ a qubit.
Multiple qubits can be used to store and process quantum information in a sufficiently redundant way so as to protect against noise.
Most of quantum error correction is concerned with such qubit codes.

Quantum information is stored in quantum superpositions of system states, e.g., superpositions of states \(|0\ket\) and \(|1\ket\) for the case of a single qubit.
Multi-qubit states are labeled by binary strings, and the message \(\rho\) to be transmitted is in general a positive-semidefinite operator on the vector space of such states.

A qubit encoding\zoo{qubits_into_qubits} maps logical quantum information redundantly into the \df{logical} or \df{code space} --- a subspace of the Hilbert space of several \df{physical qubits}.
A basis for the code space is spanned by basis elements called \df{logical codewords}.
A simple qubit encoding is the repetition code, mapping a single logical qubit into three physical qubits as \(|\mu\ket\to|\mu_L\ket=|\mu,\mu,\mu\ket\) for \(\mu\in\{0,1\}\).

An error can map the codespace into an \df{error space} spanned by basis elements called \df{error words}.
Determining which errors can be detected and corrected with a particular code can be done using the Knill-Laflamme \df{error-correction conditions}~\cite{Knill1997}\zoo{qecc_finite}.
A code is said to have \df{distance} \(d\) if errors occurring on at most \(d-1\) physical qubits can be detected.

The quantum superpositions storing logical information are susceptible not only to bit-flip errors \(X=\left(\begin{smallmatrix}0 & 1\\
1 & 0
\end{smallmatrix}\right)\) that permute the bitstring-labeled computational qubit states, but also phase-flip errors \(Z=\left(\begin{smallmatrix}1 & 0\\
0 & -1
\end{smallmatrix}\right)\) that modify the relative amplitude and phase of the superposition.
One also has to worry about combinations of the two, \(Y=\left(\begin{smallmatrix}0 & -i\\
i & 0
\end{smallmatrix}\right)\), which are often assumed to occur at a similar rate as individual bit- and phase-flips.
Multi-qubit errors are typically represented as tensor products of Pauli matrices \(X,Y,Z\) and the identity matrix \(I=\left(\begin{smallmatrix}1 & 0\\
0 & 1
\end{smallmatrix}\right)\).
This is in part because Pauli matrices conveniently form a basis for the set of all error operators, meaning that any error can be expressed as a linear combination of tensor-product \df{Pauli strings}.

In order to detect and correct errors, one performs a \df{round} (\textit{a.k.a.}\@ cycle) of error correction.
Detection is made possible by
a set of ``optimally coarse-grained'' observables called \df{check operators}, whose measurement distinguishes error spaces from the codespace without collapsing logical quantum superpositions within the respective spaces.
One round consists of (A) detecting the error via quantum measurement of check-operator eigenvalues,
which are usually called \df{syndromes}; and (B) \df{decoding}, i.e., applying a correction operation conditional on the syndrome that maps the state in the identified error space back into the codespace.

Check operators that are simple to express and measure are generally difficult to determine for a given code.
Qubit stabilizer codes~\cite{gottesman_thesis,Calderbank1996}\zoo{qubit_stabilizer} are a large class of codes whose codespaces are joint \(+1\)-eigenvalue eigenspaces of a commuting set of Pauli operators called \df{stabilizers}.
This set forms a group, and each element of this group can be represented as a product of powers of a set of generating elements.
Such \df{stabilizer generators} are precisely the check operators of a stabilizer code.
Detectable errors fail to commute with at least one stabilizer generator.
A stabilizer code encoding \(k\) logical qubits into \(n\) physical qubits and having distance \(d\) is denoted as an \([[n,k,d]]\) code.

The four-qubit code is a simple qubit stabilizer code with codewords
\begin{subequations}\label{eq:Leung}
\begin{align}
|0_{L}\ket & =\left(|0000\ket+|1111\ket\right)/\sqrt{2}\\
|1_{L}\ket & =\left(|0011\ket+|1100\ket\right)/\sqrt{2} \quad\quad\text{(four-qubit code~\cite{Vaidman1996}\zoo{stab_4_2_2}).}
\end{align}
\end{subequations}
Each codeword is \textit{itself} a quantum superposition of two canonical qubit states.
This extra redundancy helps with phase-flip noise, while the fact that the bitstrings labeling the qubit basis states present in each codeword differ in at least two positions (e.g., \(0000\) and \(0011\) differ at position 3 and 4) implies some protection against bit-flip errors.
This codespace is the joint \(+1\)-eigenvalue eigenspace of the stabilizer group
\begin{equation}\label{eq:four-qubit}
\mathsf{S}_{\text{four-qubit}}=\bra ZZII,IIZZ,XXXX \ket =\bra S_1,S_2,S_3 \ket = \{S_1^a S_2^b S_3^c~|~a,b,c\in\Z_2\}~.
\end{equation}
The stabilizer generators (\textit{a.k.a.}\@ check operators) for this code are the Pauli strings \(ZZII\), \(IIZZ\), and \(XXXX\), where, e.g., \(ZZII=Z\otimes Z\otimes I \otimes I\) acts nontrivially on qubits one and two.

\section{Bosonic modes}\label{sec:modes}

The formalism of bosonic coding begins by quantizing the electromagnetic field~\cite{scully_book} and forming bosonic modes --- quantum systems described by continuous conjugate variables (\textit{a.k.a.}\@ quadratures).
A bosonic or continuous-variable (CV) mode (\textit{a.k.a.}\@ a qumode or, more colloquially, a harmonic oscillator) can correspond to an \textit{electromagnetic} mode of a microwave cavity, an optical cavity/fiber, or free space.
Alternatively, bosonic modes can be \textit{mechanical} (\textit{a.k.a.}\@ motional or phononic), representing vibrating atomic or molecular nuclei or acoustic or nano-mechanical resonators.

While qubit canonical or computational basis states are labeled by binary strings, there exists a somewhat analogous set of canonical single-mode ``states'' \(|x\ket\) labeled by the real number \(x\in\R\), corresponding to the configuration space of a particle on a real line.
The word basis is in quotes because such \df{position ``states''} are orthogonal, but not normalizable: \(\bra x|y\ket = \delta(x-y)\), where \(\delta\) is the Dirac delta function.
Despite this, they retain several favorable and extensively utilized features:

\begin{enumerate}
	\item Position states are ``complete'', i.e., they can be used to express vectors \(|f\ket\) within the mode's Hilbert space as \begin{equation}\label{eq:pos-basis}
		|f\ket=\int_{\R} \der x |x\ket\bra x|f\ket \equiv \int_{\R} \der x f(x)|x\ket~.
	\end{equation}
	This is analogous to expressing an arbitrary qubit state in the computational basis.
	\item The position ``basis'' provides a set of generalized eigenstates with which one can diagonalize the oscillator \df{position operator} (\(\xx=\int_{\R} \der x |x\ket x \bra x|\)).
	A dual (i.e., Fourier-transformed) non-normalizable set of \df{momentum ``states''}, \(\int_\R \der x e^{i p x}|x\ket\), form a bosonic analogue of the qubit \(|0\ket\pm|1\ket\) basis and diagonalize the oscillator's \df{momentum operator} (\(\pp=-i \frac{\der}{\der x}\)).
	\item Completeness of the position ``basis'' implies that the probability \(|\bra x|f\ket|^2\) can be sampled with a bosonic \df{quantum measurement}.
	Similar continuous-parameter measurements can be formulated for momentum states and other complete sets of non-normalizable ``states''~\cite{Artiles2005,Lvovsky2008,Holevo2011}.
\end{enumerate}

Position and momentum operators satisfy the famous commutation relation \([\xx,\pp]=i\).
Realizing such a commutation relation requires an intrinsically infinite-dimensional space.
To see this, try taking the trace of both sides of the commutation relation.
Cyclic permutation of \(\xx\) and \(\pp\) under the trace, a well-defined operation for finite-dimensional spaces, yields zero for the left-hand side and thus contradicts the right-hand side.

A bona-fide orthonormal basis for a single mode consists of the \df{Fock states} \(\{|n\rangle\}_{n=0}^{\infty}\) --- eigenstates of the \df{occupation-number operator} \(\nh = \frac{1}{2}(\pp^2+\xx^2-1)\) (\textit{a.k.a.}\@ the harmonic oscillator Hamiltonian).
This Hamiltonian can be conveniently written in terms of the mode's \df{lowering operator} \(\al=\frac{1}{\sqrt{2}}(\xx+i\pp)\) and its conjugate as
\(\al^\dg\al=\nh\), and Fock states satisfy the eigenvalue relation \(\nh|n\ket=n|n\ket\).
Such states provide an alternative to Eq. (\ref{eq:pos-basis}), expanding vectors as
\begin{equation}\label{eq:fock-states}
	|f\ket = \sum_{n\geq 0} |n\ket\bra n|f\ket \equiv \sum_{n\geq 0} f_n |n\ket~.
\end{equation}

The expectation value of the oscillator Hamiltonian, \(\bra \nh \ket = \bra f | \nh | f \ket\), corresponds to the \df{average energy} of the mode in terms of the number of bosonic energy carriers, often called photons or phonons.
Similar \df{occupation-number moments} \(\bra \nh^{\ell} \ket\) are defined for higher powers \(\ell\geq 1\).

Normalizable vectors \(|f\ket\)
may admit non-normalizable ``excitations'' \(\nh^{\ell}|f\ket\) --- a subtlety not present in finite dimensions.
This is equivalent to the fact that normalizable vectors may have infinite occupation-number moments.
For example, a vector with Fock-state coefficients \(f_n = 1/\sqrt{n^{1+\epsilon}}\) is normalizable, but has infinite average energy for any \(\epsilon\in(0,1]\).
I define \df{bosonic pure states} as those vectors whose moments are finite \textit{for all} \(\ell\); the space of such vectors is called \df{Schwartz space} \footnote{Schwartz space \(S(\R)\)~\cite{Gieres2000,Madrid2005}\cite[Appx. C.3]{Iosue22}
is a subspace of the space of normalizable vectors, \(S(\R)\subset L^2(\R)\).
The position ``states'' \(\bra x|\) live in the space dual to Schwartz space, the space of tempered distributions \(S(\R)^{\prime}\supset L^2(\R)\). Together, the three spaces make up the rigged Hilbert space construction.}.
This definition is both physically meaningful and mathematically consistent, as position-``state'' overlaps \(f(x) = \bra x |f\ket\)
may be infinite for normalizable vectors not in Schwartz space, thereby ``breaking'' any position-``state'' quantum measurements.
In other words, while each normalizable vector \(|f\ket\) is in one-to-one correspondence with a square-integrable function \(f(x)\) on the real line per Eq.~(\ref{eq:pos-basis}), not all such functions
ought to be regarded as reasonable quantum states.

While I maintain that we should define our state space such that \(\nh^{\ell}|f\ket\) remains a physical quantum state, it is interesting to note that the excitation operator \(\nh^\ell\) cannot be applied on a vector as a result of any quantum channel.
Quantum channels --- the most general maps on quantum states --- consist of \df{Kraus operators} \(K\) such that the sum \(\sum_K K^\dg K\) has all eigenvalues \(\leq1\) ~\cite{nielsen_chuang}.
Since the eigenvalues of \(\nh\) grow without bound, \(\nh\) or its powers cannot be Kraus operators.
Kraus operators for common bosonic noise channels (see Sec.~\ref{sec:physical-noise}) turn out to be much tamer, belonging to the set of \df{trace-class} operators, for which \(\text{Tr}(\sqrt{K^\dg K})\) is finite.

\section{Bosonic stabilizer codes~\cite{Barnes2004}\zoo{oscillator_stabilizer}}
\label{sec:bosonic-stab}

Oscillator generalizations of the unitary Pauli bit and phase flips are represented by \df{displacements} or \df{shifts}.
For a single mode, the two types of displacements --- position shifts and momentum shifts --- act on position states as \(e^{-iq\pp}\left|y\right\rangle =\left|y+q\right\rangle\) and \(e^{iq\xx}\left|y\right\rangle =e^{iq y}\left|y\right\rangle\),
where \(q\) is any real number.
The opposite relations hold for the momentum states: \(e^{iq\xx}\) shifts the momentum while \(e^{-iq\pp}\) is diagonal, e.g.,
\begin{equation}
  e^{-iq\pp}\left(\int_{\R}\der xe^{ipx}|x\ket\right)=\int_{\R}\der xe^{ipx}|x+q\ket=e^{-iqp}\left(\int_{\R}\der xe^{ipx}|x\ket\right)~.
\end{equation}
Trace-class operators acting on multiple modes can be expanded in terms of tensor products of single-mode displacements~\cite{Cahill1969}, akin to Pauli strings in the qubit case.

Commuting sets of displacement operators define bosonic stabilizer codes in a natural extension of the qubit stabilizer formalism to bosonic modes.
One can define bosonic stabilizer codes for a single mode or multiple modes.
Groups of displacements can be continuous, yielding analog stabilizer codes, or discrete, yielding either GKP or GKP-stabilizer codes.
The logical subspace can be finite- or infinite-dimensional, depending on which group is considered.
Ideal codewords are not normalizable, and recipes for creating approximate states (lying in Schwartz space) are postponed until Sec.~\ref{sec:norm}.

\subsection{Analog stabilizer codes~\cite{Lloyd1998,Braunstein1998,Barnes2004}\zoo{analog_stabilizer}}

Position states yield an immediate generalization of the qubit repetition encoding~\cite{Peres1985}\zoo{quantum_repetition}, \(|x\ket\to|x,x,x\ket\).
This code encodes one \df{logical mode} into three physical modes, protecting from errors that cause shifts in the position of any one of the physical modes. For correcting errors that induce relative phases between position states, a dual momentum-state repetition code can be used.

Concatenating two-mode versions of the two repetition codes yields the four-mode code, an analogue of the four-qubit code (\ref{eq:Leung}), with logical position states
\begin{equation}\label{eq:four-mode}
  |x_{\text{four-mode}}\ket=\left(\int_{\R}\der ye^{ixy}|y,y\ket\right)^{\otimes2}=\int_{\R}\der y\int_{\R}\der z\,e^{ix\left(y+z\right)}|y,y,z,z\ket~.
\end{equation}
The above four-mode code detects any error on a single mode, be it a shift in position or momentum.

Notice how the four-mode codewords are \(+1\)-eigenvalue eigenvectors of the displacements \(S_1=e^{i(\xx_{1}-\xx_{2})}\), \(S_2=e^{i(\xx_{3}-\xx_{4})}\), and \(S_3=e^{-i(\pp_{1}+\pp_{2}-\pp_{3}-\pp_{4})}\). Products of powers of these three elements constitute the stabilizer group of the four-mode code,
\begin{equation}\label{eq:stab2}
  \mathsf{S}_{\text{four-mode}}= \{S_1^a S_2^b S_3^c~|~a,b,c\in\R\}~,
\end{equation}
whose joint \(+1\)-eigenvalue eigenspace is exactly the codespace.
The above group is a continuous analogue of the four-qubit code's stabilizer group (\ref{eq:four-qubit}).

All single-mode displacements do not commute with a power of at least one \(S_i\), meaning that the distance of the four-mode code (\ref{eq:four-mode}) is two. The code encodes one logical mode into four, corresponding to a \([[4,1,2]]_\R\) \df{analog stabilizer} encoding~\cite{Lloyd1998,Braunstein1998,Barnes2004}\zoo{analog_stabilizer}.
Important differences from the qubit case (\ref{eq:four-qubit}) include an infinite-dimensional logical subspace, and a continuous set of error spaces labeled by real-valued powers \(a,b,c\) of the elements \(S_i\).

Since any real power \(q\) of the \(S_i\) yields a distinct group element, it is convenient to instead study the \df{nullifiers} --- the set of mutually commuting operators in the exponentials of the elements \(S_i\) that form the stabilizer group's Lie algebra. For the four-mode code, the nullifiers are \(\xx_1-\xx_2\), \(\xx_3-\xx_4\), and \(\pp_{1}+\pp_{2}-\pp_{3}-\pp_{4}\). Since the codewords are \(+1\)-eigenvalue eigenvectors of \textit{any} power \(q\) of the stabilizer generators, they must be \(0\)-eigenvalue eigenvectors of the nullifiers. For example,
\begin{equation}
  (\xx_1-\xx_2)|x_{\text{four-mode}}\ket=\int_{\R}\der y\int_{\R}\der z\,e^{ix\left(y+z\right)}(y-y)|y,y,z,z\ket=0~.
\end{equation}
Any error that does not commute with one of the nullifiers is detectable, and the Hermitian nullifiers act as check operators that can be measured directly using established quantum measurements~\cite{Lvovsky2008}.

General \([[n,k,d]]_\R\) analog stabilizer codes encode \(k\) logical modes into \(n\) physical modes, detect errors acting on at most \(d-1\) of the modes, and correct errors acting on at most \(\lfloor\frac{d-1}{2}\rfloor\) modes. There exist analog versions of Shor's nine-qubit code~\cite{Lloyd1998}\zoo{lloyd_slotine}, realized using optical beams~\cite{Aoki2009}, the five-qubit perfect code~\cite{Braunstein1998}\zoo{braunstein}, and the surface code~\cite{Zhang2008}\zoo{analog_surface}.
Analog encodings also yield bosonic analogues of qubit cluster states and are relevant to measurement-based photonic quantum computation~\cite{Menicucci2006,Gu2009a}.

\subsection{GKP codes~\cite{Gottesman2001}\zoo{gkp}}

The four-mode code's stabilizer group (\ref{eq:stab2}) is continuous, admitting an \textit{infinite}-dimensional logical space protected against arbitrarily large displacements acting on only \textit{a few} modes. Discrete stabilizer groups are also possible, admitting a \textit{finite}-dimensional logical space protected against sufficiently small displacements acting on \textit{all} modes.

A discrete stabilizer group of commuting displacements can exist even for one mode. While displacements do not generally commute,
\begin{equation}\label{eq:commutator}
e^{-iq_1\pp}e^{iq_2\xx}=e^{iq_1 q_2}e^{iq_2\xx}e^{-iq_1\pp}\quad\quad\text{(position-momentum group commutator)},
\end{equation}
they \textit{can} commute for fine-tuned values of the displacement magnitudes \(q_{1,2}\). Consider the case \(q_1=q_2=\sqrt{2\pi N}\) for some natural number \(N\), for which the commutator phase reduces as \(e^{iq_1 q_2}=e^{i2\pi N}=1\). The displacements \(S_1=e^{i\sqrt{2\pi N}\xx}\) and \(S_2=e^{-i\sqrt{2\pi N}\pp}\) thus commute, forming a stabilizer group of the GKP code,
\begin{equation}\label{eq:gkpstab}
  \mathsf{S}_{\text{GKP}}=\bra S_{1},S_{2}\ket=\bra e^{i\sqrt{2\pi N}\hat{x}},e^{-i\sqrt{2\pi N}\hat{p}}\ket=\{S_{1}^{a}S_{2}^{b}\,|\,a,b\in\Z\}~.
\end{equation}

The above stabilizers admit a joint \(N\)-dimensional qudit \(+1\)-eigenvalue eigenspace, with qudit logical operators \(Z_L=S_1^{1/N}\) and \(X_L=S_2^{1/N}\). This eigenspace is spanned by codewords that are invariant under the discrete phase-space shifts induced by the stabilizers. Therefore, such states can be expressed as a particular \textit{lattice} of position states. For the logical-qubit case (\(N=2\)), the codewords are
\begin{equation}\label{eq:gkpstates}
  |0_{\text{GKP}}\ket=\sum_{\ell\in\Z}|\left(2\ell\right)\sqrt{\pi}\ket\quad\quad\text{and}\quad\quad|1_{\text{GKP}}\ket=\sum_{\ell\in\Z}|\left(2\ell+1\right)\sqrt{\pi}\ket\,,
\end{equation}
consisting of position states labeled by points on two alternating 1D lattices whose points are spaced apart by \(2\sqrt{\pi}\). Each position state is a \(+1\)-eigenvalue eigenvector of the stabilizer \(S_1\), and the lattice spacing ensures that both lattices are invariant under position shifts (by discrete multiples of \(2\sqrt{\pi}\)) induced by \(S_2\) and its powers.

\begin{figure}
\centering
\includegraphics[width=0.75\textwidth]{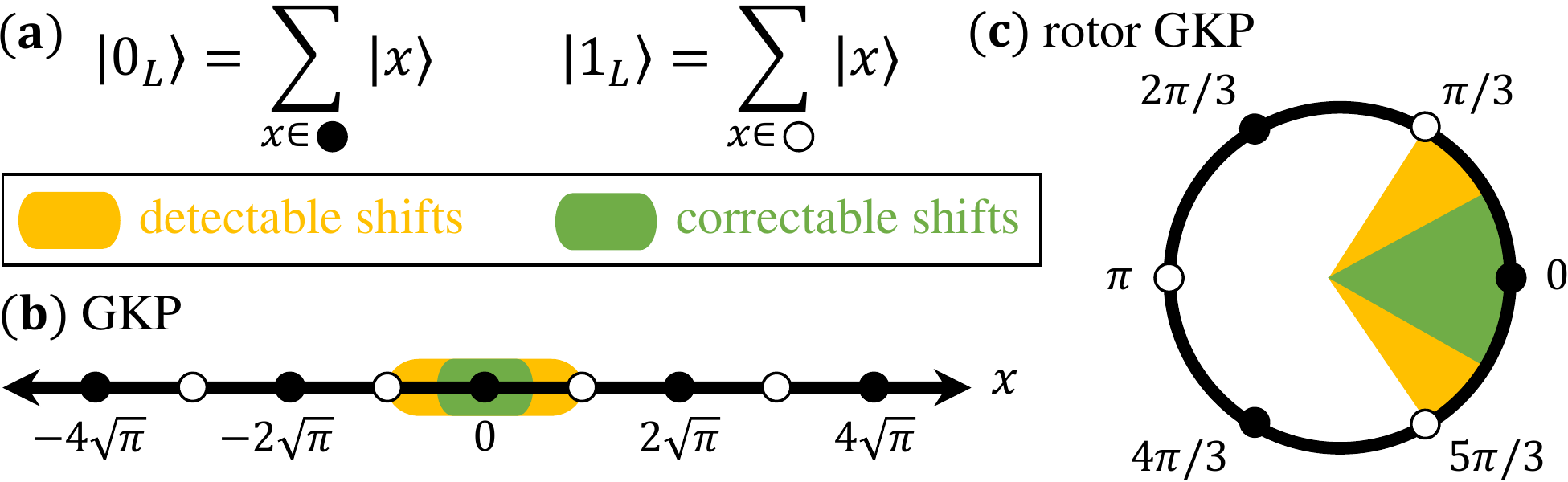}    
\caption{{\bf (a)} GKP codewords are superpositions of alternating position states \(x\) and can be similarly formulated for linear and angular systems, i.e., bosonic modes and planar rotors. I sketch position states participating in {\bf (b)} square-lattice GKP codewords (\ref{eq:gkpstates}) and {\bf (c)} \(N=3\) rotor-GKP codewords as well as regions of detectable (yellow) and correctable (green) shifts.}
\label{f:gkp}
\end{figure}

The above code detects any position shifts that shift the logical states out of the codespace. The smallest distance between position states on opposite codewords is \(\sqrt{\pi}\), so shifts up to that amount in either direction are detectable. The detectable-shift magnitudes \(q\in(-\sqrt{\pi},\sqrt{\pi})\) correspond to the unit cell of the lattice formed by the logical-zero position-state labels [black circles in Fig.~\ref{f:gkp}(b)].

While the above argument applies to position shifts only, GKP codes can also be represented as a comb of evenly spaces momentum states. This yields an almost identical argument for protecting against momentum shifts by \(p\in(-\sqrt{\pi},\sqrt{\pi})\).

There are no nullifiers for this code since the stabilizer group is discrete.
Detectable displacements fail to commute with \(S_1\) and/or \(S_2\), and the continuum of error spaces is labeled by eigenvalues of the two check operators. Measuring their eigenvalues determines the mode's position and momentum modulo \(\sqrt{\pi}\). In the lattice analogy, this determines the location of the error word in the unit cell \((-\sqrt{\pi}/2,\sqrt{\pi}/2)\) of the lattice formed by \textit{both} the logical-zero and logical-one position-state labels [black and white circles in Fig.~\ref{f:gkp}(b)]. This unit cell is in one-to-one correspondence with the set of correctable displacements.

The GKP stabilizer generators \(S_1,S_2\) can be thought of as generators of translations of a 2D square lattice in the mode's phase space. Generalizations of the square-lattice GKP code include rectangular-lattice GKP codes protecting against larger displacements in one quadrature at the expense of lesser protection against displacements in the other quadrature. Extending to codes on \(n\) modes, the stabilizer group generated by a pair of square-lattice GKP stabilizers for each mode yields a multimode GKP code, corresponding to a \(2n\)-dimensional hypercubic lattice. Still other lattices can be used to construct more exotic multimode GKP codes~\cite{Gottesman2001,Harrington2001}\zoo{multimodegkp} with unit cells that correspond to different sets of correctable displacements.

\subsection{GKP-stabilizer codes~\cite{Noh2019a}\zoo{gkp-stabilizer}}
\label{sec:gkp-stab}

GKP codes on \(n\) modes with \(2n\) stabilizer generators support a finite-dimensional logical subspace,
but removing some of the GKP stabilizers makes the logical dimension infinite while maintaining some of the protection.
These codes are the most recent addition to the bosonic stabilizer class, so I devote more space to describe their abilities.

Consider a two-mode GKP stabilizer group generated by square-lattice generators (\ref{eq:gkpstab}) with \(N=1\) supported on mode two. The joint \(+1\)-eigenvalue eigenspace of such a lopsided group consists of any state on the first mode (since there are no stabilizers supported there) tensored with the ``canonical'' GKP state on the second mode,
\begin{equation}\label{eq:gkpcomb}
  |\textsc{GKP}\ket=\sum_{\ell\in\Z}|\ell\sqrt{2\pi}\ket, \quad\quad\text{with stabilizer generators}\quad\quad e^{i\sqrt{2\pi}\xx_2}\quad\text{and}\quad e^{-i\sqrt{2\pi}\pp_2}~.
\end{equation}

To suppress shift errors on the first mode, we apply an entangling operation \textit{before and after} sending the code though a displacement-noise channel. The \(\textsc{csum}\) gate transforms the four quadratures \(\{\xx_{1},\pp_{1},\xx_{2},\pp_{2}\}\) as \(\xx_2\to\xx_2+\xx_1\), \(\pp_1\to\pp_1-\pp_2\), and leaves the other two intact. The inverse of \(\textsc{csum}\) is applied after the displacement noise. After the complete \(\textsc{csum}\to\textsc{noise}\to\textsc{csum}^{\dagger}\) sequence, the four quadratures transform as
\begin{equation}
\label{eq:quads}
\begin{array}{ccc}
\begin{alignedat}{1}
  \xx_1 & \to \xx_1 + \delta x_1\\
  \pp_1 & \to \pp_1 + \delta p_1 + \delta p_2
\end{alignedat}
& \,\,\,\,\,\,\,\,\,\,\,\,\,\,\,\,\,\,\,\,\,\,\,\,\,\,\,\,\,\,\,\,\,\, & \begin{alignedat}{1}
  \xx_2 & \to \xx_2 + \delta x_2 - \delta x_1\\
  \pp_2 & \to \pp_2 + \delta p_2
\end{alignedat}
\end{array}\,,
\end{equation}
where \(\{\delta x_{1},\delta p_{1},\delta x_{2},\delta p_{2}\}\) are small random shifts induced by the noise channel.

Information about the first mode's displacement errors that was transferred to the second mode via the \(\textsc{csum}\) gate can now be extracted via stabilizer measurements. Measuring the code's stabilizer generators (\ref{eq:gkpcomb}) extracts the second mode's displacement errors, yielding outcomes \(x^{\text{obs}}=\delta x_{2}-\delta x_{1}\) and \(p^{\text{obs}}=\delta p_{2}\). Both errors are measured modulo \(\sqrt{2\pi}\) --- the spacing of that mode's GKP state (\ref{eq:gkpcomb}) --- but this can be disregarded for simplicity by assuming that both shifts are sufficiently small.

To recover against the momentum shift in the first mode, a momentum displacement by \(-p^{\text{obs}}=-\delta p_2\) is applied, thereby correcting the part of the shift from Eq. (\ref{eq:quads}) that came from mode two, \(\pp_1 + \delta p_1 + \delta p_2\to \pp_1 + \delta p_1\).

On the other hand, exact correction of the position shift \(\delta x_1\) is not possible because its value cannot be obtained from the measurement outcome \(x^{\text{obs}}\).
Instead, one obtains a maximum-likelihood estimate \(\delta x_1^{\text{est}}\) of the true shift error using knowledge of \(x^{\text{obs}}\) and the assumption that the most likely overall shift magnitude is the smallest one,
\begin{equation}
  \delta x_{1}^{\text{est}}=\argmin_{\delta x_{1}}~(\delta x_{1})^{2}+(\delta x_{2})^{2}=\argmin_{\delta x_{1}}~(\delta x_{1})^{2}+(x^{\text{obs}}+\delta x_{1})^{2}=-\frac{x^{\text{obs}}}{2}\,.
\end{equation}
Displacing the the first mode's position back by the above amount transforms the first position quadrature from Eq. (\ref{eq:quads}) as \(\xx_{1}+\delta x_{1}\to\xx_{1}+\delta x_{1}+\frac{x^{\text{obs}}}{2}=\xx_{1}+\frac{1}{2}(\delta x_{1}+\delta x_{2})\).

In summary, the combination of \(\textsc{csum}\) conjugation and GKP recovery yields output displacement errors \(\delta p_1\) and \(\frac{1}{2}(\delta x_{1}+\delta x_{2})\) on the position and momentum quadratures of the logical mode, respectively. How is this an improvement?
For simplicity, let's assume that the four input shift errors \(\{\delta x_{1},\delta p_{1},\delta x_{2},\delta p_{2}\}\) coming from a displacement-noise channel are independent and distributed according to a Gaussian distribution with mean zero and variance \(\sigma^2\).
Due to the tendency of sums of such random variables to converge to their mean, the sum \(\frac{1}{2}(\delta x_{1}+\delta x_{2})\) follows a distribution with variance \(\sigma^2/2\), \textit{half} as much as that of the noise \(\delta x_1\) in an unencoded mode.

The above GKP-repetition~\cite{Noh2019a}\zoo{gkp-stabilizer} encoding yields a reduction in the position noise of a logical mode \textit{without} increasing the noise in the momentum. Using more modes and other gates in addition to \(\textsc{csum}\), it is possible to reduce noise in \textit{both} quadratures in similar fashion. Other GKP-stabilizer codes can
suppress the output variance by some power (up to constants and logarithmic corrections), e.g., \(\sigma^{2}\to\sigma^{4}\) for the GKP-two-mode-squeezing code. The above maximum-likelihood decoder holds for general versions of such codes~\cite{Xu2022}.

\subsection{Normalization and precision}\label{sec:norm}

The bosonic stabilizer codewords that we have discussed so far are unphysical.
Logical position states of the analog repetition encoding, \(|x_L\ket=|x,x,x\ket\), overlap as \(\bra x_L|y_L\ket=[\delta(x-y)]^3\), which is not the proper \(\delta\)-function normalization condition of single-more position states. GKP states span a finite-dimensional logical subspace, yet are also not normalizable. As a result, error correction as prescribed above will not be possible for such codes in the real world, and approximations have to be made.

Approximate codewords can be created by substituting each position state \(|x\ket\) in the ideal codewords as a Gaussian of width \(\Delta\) centered at \(x\). This is equivalent to applying a linear combination of Gaussian-weighed position displacements to the ideal position state (up to a normalization constant),
\begin{equation}
  |x\ket\to|x^{(\Delta)}\ket\propto e^{-\frac{1}{2}\Delta^{2}\pp^{2}}|x\ket={\textstyle \frac{1}{\sqrt{2}}}\int_{\R}{\textstyle \der y}e^{-\frac{1}{2}y^{2}/\Delta^{2}}e^{-iy\pp}|x\ket={\textstyle \frac{1}{\sqrt{2}}}\int_{\R}\der ye^{-\frac{1}{2}(y-x)^{2}/\Delta^{2}}|y\ket~.
\end{equation}
The first equality is obtained by treating \(\pp\) as a real number (by temporarily inserting a resolution of identity in terms of momentum states) and expressing the exponential in terms of its Fourier transform.
The second equality results from applying the displacement and changing variables in the integral.
The new state can be thought of as squeezed in the position quadrature, reducing to the infinitely squeezed position state as \(\Delta\to 0\).

In case of GKP codewords, a dual position-state envelope \(e^{-\frac{1}{2}\Delta^{2}\xx^{2}}\) is also required because there is an infinite number of position states present in each codeword. Up to a normalization constant, this yields
\begin{equation}
  |\textsc{GKP}\ket\to|\textsc{GKP}^{(\Delta)}\ket\propto e^{-\frac{1}{2}\Delta^{2}\pp^{2}}e^{-\frac{1}{2}\Delta^{2}\xx^{2}}\sum_{\ell\in\Z}|\ell\sqrt{2\pi}\ket=\sum_{\ell\in\Z}e^{-\pi\Delta^{2}\ell^{2}}|\ell\sqrt{2\pi}\,^{(\Delta)}\ket
\end{equation}
for the canonical GKP state (\ref{eq:gkpcomb}). The joint effect of momentum- and position-state envelopes can be implemented with a \df{cooling operator} (\textit{a.k.a.}\@ damping operator or regularizer)~\cite[Appx. B]{Menicucci2014},
\begin{equation}\label{eq:cooling}
	e^{-\beta\nh}=e^{-\frac{1}{2}\beta\left(\pp^{2}+\xx^{2}-1\right)}\quad\quad\text{for}\quad\quad\beta\in(0,\infty)~.
\end{equation}

Cooled GKP codewords yield an intrinsic probability of misidentification
due to non-orthogonality of the codewords, but such an error is exponentially suppressed as \(\Delta \to 0\). Displacement noise interacts very simply with cooled GKP states, merely increasing the sizes of the position and momentum envelopes.

Another complication of bosonic stabilizer codes is that it is impossible to keep track of a continuum of error spaces due to the finite precision of any measurement apparatus,
which leads to additional imperfections~\cite{Terhal2016}.
Coarse-graining this continuum into a finite number \(M\) of possible outcomes requires an ancillary system of at least dimension \(M\),
presenting overhead not present in the qubit case, in which measurements are typically binary.
Remarkably, recent experiments~\cite{CampagneIbarcq2019,DeNeeve2020} (for theory, see~\cite[Supplement]{Royer2020}) have performed beneficial GKP error correction by extracting syndromes to only \textit{one bit} of precision, i.e., requiring the smallest possible overhead of \(M=2\).

\section{Bosonic Fock-state codes\zoo{fock_state}}\label{sec:fock-state}

Extending the principles of QEC to oscillators using position states yields bosonic stabilizer codes, but a different set of codes emerges when the Fock state is the fundamental ingredient.

\subsection{Number-phase codes~\cite{Grimsmo2019}\zoo{number_phase}}

GKP codes are based on lattices in (linear) position and momentum. Analogues of GKP codes for the rotational degrees of freedom of a \df{planar rotor}~\cite{Albert2017} are called rotor GKP codes~\cite{Gottesman2001,mol}\zoo{rotor_gkp}.

The canonical set of states for a planar rotor are its position ``states'' \(\{|\phi\ket_{\text{rot}}~,~\phi\in[-\pi,\pi)\}\), whose label corresponds to the position of a particle on a ring (as opposed to a particle on a line in the case of a bosonic mode).
Since the position degree of freedom is compact, the resulting momentum degree of freedom is disrete, with corresponding momentum states \(|\ell\ket_{\text{rot}}\) labeled by an integer \(\ell\in\Z\).
The Fourier series maps between the two sets of states, e.g., \(|\phi\ket_{\text{rot}} = \frac{1}{\sqrt{2\pi}}\sum_{\ell\in\Z}e^{i\phi\ell}|\ell\ket_{\text{rot}}\).

In an analogue of linear GKP states [Fig.~\ref{f:gkp}(a)], rotor GKP codewords consist of alternating combs of rotor position ``states'' [Fig.~\ref{f:gkp}(b)].
Such codes protect against sufficiently small rotations of the rotor's angular position as well as kicks in its angular momentum.

\begin{figure}
\centering
\includegraphics[width=0.75\textwidth]{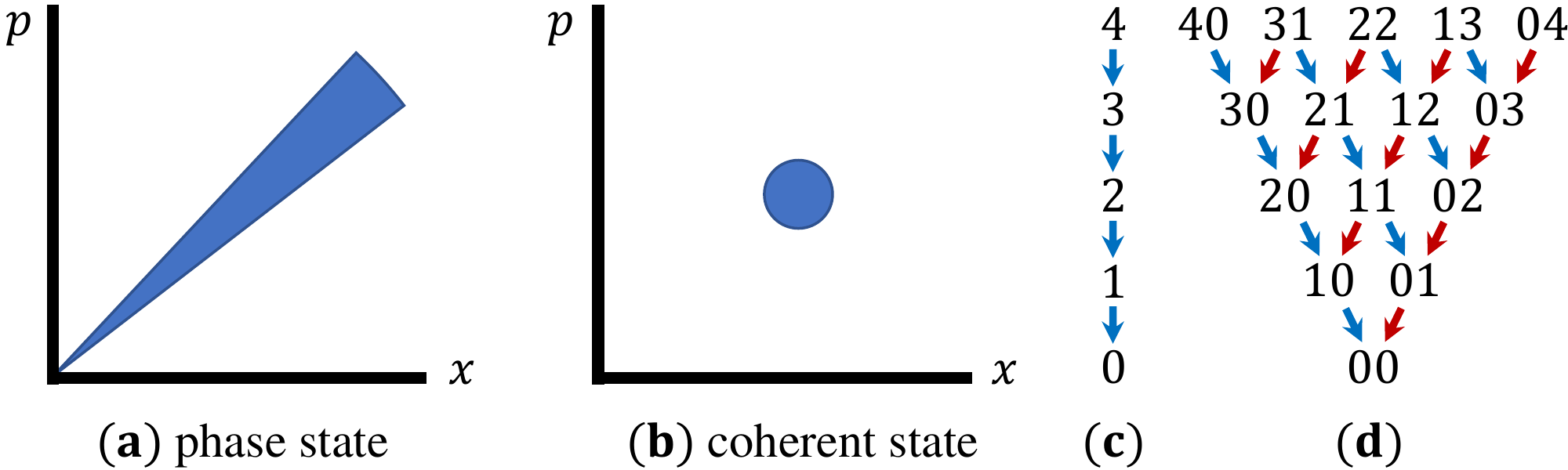}    
\caption{
Sketch of Wigner function of \textbf{(a)} a phase state and \textbf{(b)} a coherent state in the mode's position-momentum phase space.
Sketch of \textbf{(c)} single-mode and \textbf{(d)} two-mode ladders of Fock states with occupation number up to four. Loss operators \(a\) and \(b\) decrease the occupation number in either mode by one, and the Fock-state spacing provided by Fock-state codes allows for detection and correction of such errors.
}
\label{fig:fock}
\end{figure}

Poor man's versions of planar-rotor states can be embedded into a single bosonic mode.
The oscllator phase ``states'' [see Fig.~\ref{fig:fock}(a)],
\begin{equation}\label{eq:phase}
  |\phi\ket=\frac{1}{\sqrt{2\pi}}\sum_{n\geq0}e^{i\phi n}|n\ket\quad\quad\text{(phase states)}~,
\end{equation}
are bosonic analogues of angular position states.
Fock states \(|n\ket\) play the role of a ``dual'' angular-momentum basis, with the generator \(\nh\) playing the role of a pseudo-momentum operator.
The significant obstruction to a full-fledged rotor analogy is that the momentum
is never negative.
Because of this, phase ``states'' are not only non-normalizable, but they are also not quite orthogonal.
This makes for more subtle calculations, but does not qualitatively obscure the rotor interpretation.

Number-phase codes are bosonic analogues of rotor GKP codes.
Codewords can be expressed as equal superpositions of phase states at alternating ``combs'' of \(N\) angular positions; for \(N=3\) [see Fig.~\ref{f:gkp}(c)],
\begin{subequations}\label{eq:number-phase}
\begin{align}
  |0_{\text{num-ph}}\ket&=\left(\left|\phi=0\right\rangle +\left|\phi={\textstyle \frac{2\pi}{3}}\right\rangle +\left|\phi={\textstyle \frac{4\pi}{3}}\right\rangle \right)/\sqrt{3}\\
	|1_{\text{num-ph}}\ket&=\left(\left|\phi={\textstyle \frac{\pi}{3}}\right\rangle +\left|\phi=\pi\right\rangle +\left|\phi={\textstyle \frac{5\pi}{3}}\right\rangle \right)/\sqrt{3}~.
\end{align}
\end{subequations}
Such states protect against angular rotations \(e^{i\varphi\nh}\) that shift the position-like phase states as \(e^{i\varphi\nh}|\phi\ket=|\phi+\varphi\ket\). Per an argument identical to the linear GKP case, rotations with \(|\varphi|<\frac{\pi}{N}\) \textit{should be} detectable, and those with \(|\varphi|<\frac{\pi}{2N}\) \textit{should be} correctable. However, since phase states are not quite orthogonal, a finer analysis in the limit of small rotations has to be performed (see below).

An alternative basis for the number-phase codespace at \(N=3\),
\begin{subequations}\label{eq:phase-dual}
\begin{align}
  |+_{\text{num-ph}}\ket&=\left(|0_{\text{num-ph}}\ket+|1_{\text{num-ph}}\ket\right)/\sqrt{2}\propto\sum_{n\geq0}|6n\ket\\|-_{\text{num-ph}}\ket
	&=\left(|0_{\text{num-ph}}\ket-|1_{\text{num-ph}}\ket\right)/\sqrt{2}\propto\sum_{n\geq0}|6n+3\ket~,
\end{align}
\end{subequations}
reveals codewords consisting of alternating combs of Fock states. Generally, the logical-plus is supported on \(|0\ket,|2N\ket,\cdots\), while logical-minus is supported on \(|N\ket,|3N\ket,\cdots\). This yields protection against noise that changes the occupation number: kicks with magnitude at most \(N-1\) are detectable, and kicks with magnitude at most \(\lfloor(N-1)/2\rfloor\) are correctable.

While a natural candidate for a momentum-kick operator is the lowering operator \(\al\), a closer analogue is its ``regularized'' version, \(\EE_{-1}={\textstyle \frac{1}{\sqrt{\nh+1}}}\al\), cleansed of unwieldy square-root factors.
Powers of this \df{phase operator} \(\EE_{-1}\)~\cite{Susskind1964} and its adjoint (\textit{a.k.a.}\@ phasors~\cite{Bergou1991}),
\begin{equation}\label{eq:phaseop}
  \EE_{\ell}=\begin{cases}
~{\displaystyle \sum_{n\geq 0}}|n+\ell\ket\bra n| & \ell\geq0\\
~{\displaystyle \sum_{n\geq 0}}|n\ket\bra n+|\ell||=\EE_{|\ell|}^{\dg} & \ell<0
\end{cases}\quad\text{for}\quad\ell\in\Z~,
\end{equation}
form the closest bosonic analogue of rotor momentum kicks.

For \(\{\EE_\ell\}_{\ell\in\Z}\) to be bona-fide momentum kicks, they would need to raise or lower the angular momentum by \(\ell\) and to be diagonal in the position basis.
These operators fail gracefully to accomplish both tasks due to the lack of negative Fock states.
They act as momentum kicks by raising or lowering Fock states, but sometimes annihilate Fock states since there is nowhere to go below \(n=0\).
They are ``almost'' diagonal in the position basis, admitting phase states as either right or left eigenstates.
For example, \(\EE_{\ell}^\dg\) for \(\ell>0\) admit phase states (\ref{eq:phase}) as right eigenstates, analogous to \(\al^\ell\), which is also not unitary but admits coherent states as right eigenstates.
Despite such setbacks, rotations and Fock-state momentum kicks retain the proper angular commutation relation,
\begin{equation}\label{eq:num-phase-commutator}
e^{i\phi\nh}\EE_{\ell} = e^{i\phi\ell}\EE_{\ell}e^{i\phi\nh} \quad\quad\text{(number-phase commutator)}~,
\end{equation}
complementary to the linear position-momentum commutator (\ref{eq:commutator}).

Linear combinations of Fock-space rotations and the above ``momentum kicks'' can express any trace-class operator~\cite[Sec. VI.A]{Grimsmo2019}\cite{dephasing} in a \textit{polar-like} decomposition, akin to the \textit{cartesian-like} decomposition in terms of position and momentum shifts. However, due to non-orthogonality of the phase states, this set fails to be an orthonormal operator basis.
These features can be summarized by
\begin{subequations}
	\begin{align}
	\sum_{\ell\in\Z}\int_{-\pi}^{\pi}{\textstyle \frac{d\phi}{2\pi}}\langle m|\EE_{\ell}e^{i\phi\nh}|n\rangle\langle n^{\prime}|e^{-i\phi\nh}\EE_{\ell}^{\dagger}|m^{\prime}\rangle &=\delta_{nn^{\prime}}\delta_{mm^{\prime}}\quad\quad\quad\quad\text{(completeness)}\,,\\
	\text{Tr}\left(e^{-i\phi\nh}\EE_{\ell}^{\dagger}\EE_{\ell^{\prime}}e^{i\phi^{\prime}\nh}\right)=\delta_{\ell\ell^{\prime}}\bra\phi|\phi^{\prime}\ket &\neq\delta_{\ell\ell^{\prime}}\delta\left(\phi-\phi^{\prime}\right)\quad\quad\text{(non-orthogonality)}~.
	\end{align}
\end{subequations}

Rotations and kicks partially preserve the stabilizer formalism. The number-phase codespace is a \(+1\)-eigenvalue eigen-subspace of the rotation \(e^{i\frac{2\pi}{N}\nh}\).
Since phase states \(|\phi\ket\) are right eigenstates of \(\EE_{\ell}^\dg\) with eigenvalue \(e^{i\phi\ell}\), number-phase codes form the joint \(+1\) \textit{right} eigenspace of the rotation and the kick \(\EE_{2N}^\dg\). These ``stabilizer generators'' generate
\begin{equation}
	\textsf{S}_{\text{num-phase}}=\{e^{i\frac{2\pi}{N}\nh a}\EE^\dg_{2Nb}~|~a\in\Z_{N},\,b\in\mathbb{N}_{0}\}~,
\end{equation}
that fails to be a group since the kick is not unitary (\(\mathbb{N}_0\) denotes the natural numbers).
Nevertheless, the generators correspond to measurable number-phase error syndromes.

Eigenvalues of the \(\frac{2\pi}{N}\)-rotation are \(N\)th roots of unity, \(e^{i\frac{2\pi}{N}j}\) for \(j\in\Z_{N}\), whose \(j\)th eigenspace is supported by Fock states with labels \(n=j\) modulo \(N\). The \df{modular number} measurement extracting the syndrome value \(j\) can be performed using an ancilla with \(N\) possible outcomes.
Using a superconducting qubit ancilla, the occupation-number parity measurement for \(N=2\) has become a standard subroutine in microwave cavity modes ~\cite{Ofek2016,Rosenblum2018}.

Measuring in the right-eigenstate (i.e., phase-state) ``basis'' of the phase operator \(\EE_{1}^\dg\) corresponds to measuring the phase induced on a state by a rotation~\cite{Helstrom1969,Holevo2011,Grimsmo2019}. Since number-phase codewords are superpositions of phase states, the syndrome associated with the dual ``stabilizer'' \(\EE_{2N}^\dg\) is the phase \textit{modulo} \(\frac{2\pi}{2N}\) --- the \df{modular phase}. This phase can be extracted via a modular phase measurement~\cite{Grimsmo2019,dephasing} performed in the basis of number-phase code and error words.

\subsection{Cat codes~\cite{Leghtas2013b}\zoo{cat}}

Number-phase codewords (\ref{eq:number-phase}), like GKP codewords (\ref{eq:gkpstates}), are not normalizable. Substituting phase states
for \df{coherent states} [see Fig.~\ref{fig:fock}(a-b)],
\begin{equation}
  |\phi\ket\propto\sum_{n\geq 0}e^{i\phi n}|n\ket\quad\to\quad
  e^{-\frac{1}{2}\alpha^2}\sum_{n\geq0}\frac{\alpha^{n}}{\sqrt{n!}}e^{i\phi n}|n\ket\quad\quad\text{for large real \(\alpha\)}~,
\end{equation}
converts number-phase codes to cat codes --- a normalizable alternative inheriting some of the protection.

Cat codewords are superpositions of \(N\) coherent states
distributed equidistantly on a circle in the mode's phase space.
Since coherent states can be thought of as regularized phase states, the Fock-state spacing of Eq. (\ref{eq:phase-dual}) and corresponding protection against small Fock-state shifts is maintained by the codes. As a result, the rotation \(e^{i\frac{2\pi}{N}\nh}\) remains a ``stabilizer'', and measuring the modular occupation number detects such shifts.

Coherent states, like phase states, are not orthogonal, and a proper way to determine protection against rotation errors
is to examine small rotation angles. Expanding \(e^{i\phi\nh}\approx I+i\phi\nh\) and inspecting the Knill-Laflamme error-correction conditions~\cite{Knill1997}\zoo{qecc_finite}, cat codes approximately detect \df{dephasing errors} \(\nh\) up to corrections exponential in \(-\alpha^2\),
\begin{equation}\label{eq:cat-dephasing}
  P_{\text{cat}}\nh P_{\text{cat}}=\alpha^{2} P_{\text{cat}}+O(\alpha^{2}e^{-\alpha^{2}})Z_{\text{cat}}~.
\end{equation}
Above, \(P_{\text{cat}}\) is the projection onto the cat-state codespace, and \(Z_{\text{cat}}=|+_{\text{cat}}\ket\bra+_{\text{cat}}|-|-_{\text{cat}}\ket\bra-_{\text{cat}}|\).
A similar equation holds for powers \(\nh^{\ell}\). The \(Z\)-type uncorrectable term stems from unequal expectation values of \(\nh\) in the codewords, \(\bra+_{\text{cat}}|\nh|+_{\text{cat}}\ket\neq\bra -_{\text{cat}}|\nh|-_{\text{cat}}\ket\), which in turn is due to exponentially decaying overlap between different coherent states.
Besides exponential suppression of the \(Z\)-type undetectable-error term in the large-\(\alpha\) limit, cat codes also admit fine-tuned ``sweet-spot'' values of \(\alpha\) at which the term is identically zero (e.g.,~\cite{codecomp}).
Since \(\nh\) does not cause transitions between Fock states, \(X\)- and \(Y\)-type errors do not occur.

Cat codewords are not right eigenstates of the number-phase ``stabilizer'' \(\EE_{-2N}\).
The modular-phase measurement is thus not compatible with cat codes because it collapses cat codewords into number-phase states.
However, the partial phase information extracted from such a measurement can be combined with freshly initialized logical states~\cite{Knill2003,Knill2004} to yield a teleportation-based correction scheme~\cite{Grimsmo2019}.

Cat codewords \textit{are} right eigenstates of a power of the lowering operator, \(\al^{2N}\), with eigenvalue \(\alpha^{2N}\).
Thid yields a quantum-optical stabilization scheme that utilizes \df{Lindbladian} evolution, \(\frac{\partial\rho}{\partial t}=2J\rho J^{\dg}-J^{\dg}J\rho-\rho J^{\dg}J\) with jump operator \(J=\al^{2N}-\alpha^{2N}\), to drive the system into the codespace in the infinite-time limit~\cite{Wolinsky1988,Krippner1994,Hach1994,Gilles1994}.
Cat-code Lindbladian-based schemes provide ``passive'' protection against dephasing noise, and can be thought of as series of ``active'' syndrome measurements and corrections occurring instantaneously and continuously~\cite{Paz1998}.
The \textit{two-component} \(N=1\) cat code~\cite{Cochrane1999}\zoo{two-legged-cat}, originally proposed for quantum computing with optical platforms \cite{Jeong2001,Ralph2003,Lund2007}, has been ``stabilized'' this way at microwave frequencies~\cite{Leghtas2014,S.Touzard,Lescanne2019}.
The Kerr-cat qubit~\cite{Puri2017} --- a related \(N=1\) scheme utilizing the Hamiltonian \(J^\dg J\) --- has also been realized~\cite{Grimm2019}.


\subsection{Binomial codes~\cite{bin}\zoo{binomial}}

Binomial codes are the third class of single-mode codes that, along with number-phase and cat codes, share the same mechanism for
protecting against Fock-space shifts; these three code families make up the bosonic rotation codes~\cite{Grimsmo2019}\zoo{bosonic_rotation}.
The difference with binomial codes is that they are supported on only a finite set of Fock states. They are also more ``digital''
in their protection against dephasing errors, detecting powers of \(\nh\) \textit{exactly} up to a maximum power \(D\).

The simplest binomial code can be obtained from the four-qubit code (\ref{eq:four-qubit}), whose codewords are \(\frac{1}{\sqrt{2}}\left(|0000\ket+|1111\ket\right)\) and \(\frac{1}{\sqrt{2}}\left(|0011\ket+|1100\ket\right)\). Adding up each of the digits in the codewords' binary strings, \(|0000\ket\to|0\ket\), \(\{|1100\ket,|0011\ket\}\to|2\ket\), and \(|1111\ket\to|4\ket\), yields the single-mode ``0-2-4'' binomial code with codewords
\begin{equation}\label{eq:binomial}
|0_{\text{bin}}\ket = (|0\ket+|4\ket)/\sqrt{2}\quad\quad\text{and}\quad\quad|1_{\text{bin}}\ket=|2\ket~.
\end{equation}

Due to the Fock-state spacing of \(N=2\) between the codewords [see Fig.~\ref{fig:fock}(c)], the above code detects a single Fock-state shift, either up or down (but not both). Because the average occupation number \(\bra\nh\ket\) of both codewords is two, the code also detects a dephasing error \(\nh\). Together, these imply \textit{correction} of the error set \(\{I,\al\}\), i.e., correction of a single \df{loss error}. This is analogous to the four-qubit code's ability to correct a single decay event \(\sigma_-=\frac{1}{\sqrt{2}}(X-iY)\) on any physical qubit. A similar argument yields protection against a single \df{excitation} \(\al^{\dagger}\).

General binomial codes, with codewords
\begin{equation}
	|0/1_{\text{bin}}\ket={\textstyle{\frac{1}{2^{D/2}}}}\sum_{m\sim\text{even}/\text{odd}}\sqrt{\binom{D+1}{m}}|Nm\ket~,
\end{equation}
are defined by non-negative integer parameters \(N,D\geq 0\); for the ``\(0,2,4\)'' code (\ref{eq:binomial}), \(N=2\) and \(D=1\). The parameter \(N\) sets the Fock-state spacing of the code, relying on modular number measurements to detect photon losses and excitations.
The parameter \(D\) sets the coefficient in front of each Fock state to ensure that both codewords have equal moments, \(\bra 0_{\text{bin}}|\nh^{\ell}|0_{\text{bin}}\ket=\bra 1_{\text{bin}}|\nh^{\ell}|1_{\text{bin}}\ket\), for \(\ell\leq D\); this allows for detection of up to \(D\)-order dephasing errors. Correcting dephasing and loss errors simultaneously remains difficult, but some progress has been made by relating binomial codes to spin-coherent states that are staggered in Fock space~\cite{codecomp}.

\subsection{Chebyshev codes~\cite{Layden2019}\zoo{chebyshev}}

A related ``digital'' code called the Chebyshev code protects against the same dephasing errors \(\{\nh^\ell~|~\ell\leq D\}\) as the binomial codes, but optimizes the leading-order logical-\(Z\) error, \(\bra 0_{\text{cheb}}|\nh^{D+1}|0_{\text{cheb}}\ket-\bra 1_{\text{cheb}}|\nh^{D+1}|1_{\text{cheb}}\ket\),
for sensing applications.
This error also yields a logical gate \(\exp(i t \nh^{D+1})\) on the codespace.
In the context of local quantum sensing, such a gate can be interpreted as an external signal whose corresponding time \(t\) is a parameter that needs to be ``sensed'' or measured.
Chebyshev codes achieve near-optimal sensitivity to such a signal while maintaining error-correcting properties.

\subsection{Multi-mode Fock-state codes\zoo{fock_state}}

The notion of Fock-state spacing can be extended to yield related multimode Fock-state codes defined on a finite subspace of Fock states [see Fig.~\ref{fig:fock}(d)]. The simplest such code, ubiquitous in optical platforms~\cite{Knill2001}, is the two-mode dual-rail code~\cite{Chuang1995}\zoo{dual_rail} with codewords
\begin{equation}\label{eq:dual-rail}
|0_{\text{dual-rail}}\ket = |10\ket \quad\quad\text{and}\quad\quad |1_{\text{dual-rail}}\ket=|01\ket~,
\end{equation}
detecting one loss error in either mode by checking whether a photon has leaked out.
The two modes of the encoding can represent spatial or temporal modes, corresponding, respectively, to a \textit{frequency-bin} or \textit{time-bin} encoding.

The more complex Chuang-Leung-Yamamoto (CLY) codes are based on a generalized Hamming distance in multimode Fock space and protect against loss errors in multiple modes~\cite[Eq. (4.11)]{Chuang1997}\zoo{chuang-leung-yamamoto}.
A two-mode CLY code admits codewords \(\frac{1}{\sqrt{2}}(|04\ket+|40\ket)\) and \(|22\ket\), utilizing Fock-state spacing within both modes. Similar properties holds for a three-mode code~\cite{Wasilewski2007}\zoo{wasilewski-banaszek} with codewords \(\frac{1}{\sqrt{3}}(|300\ket+|030\ket+|003\ket)\) and \(|111\ket\).
Both codes are permutation invariant and  contain the same number of total excitations in each participating Fock state; the latter property makes them immune to any errors that are functions of the total occupation number \(\sum_i \nh_i\).
This immunity and permutation invariance delineate a special family of CLY codes~\cite{Ouyang2018}\zoo{constant_excitation_permutation_invariant}.

\section{The fight against physical noise}
\label{sec:physical-noise}

\begin{figure}
\centering
\includegraphics[width=0.85\textwidth]{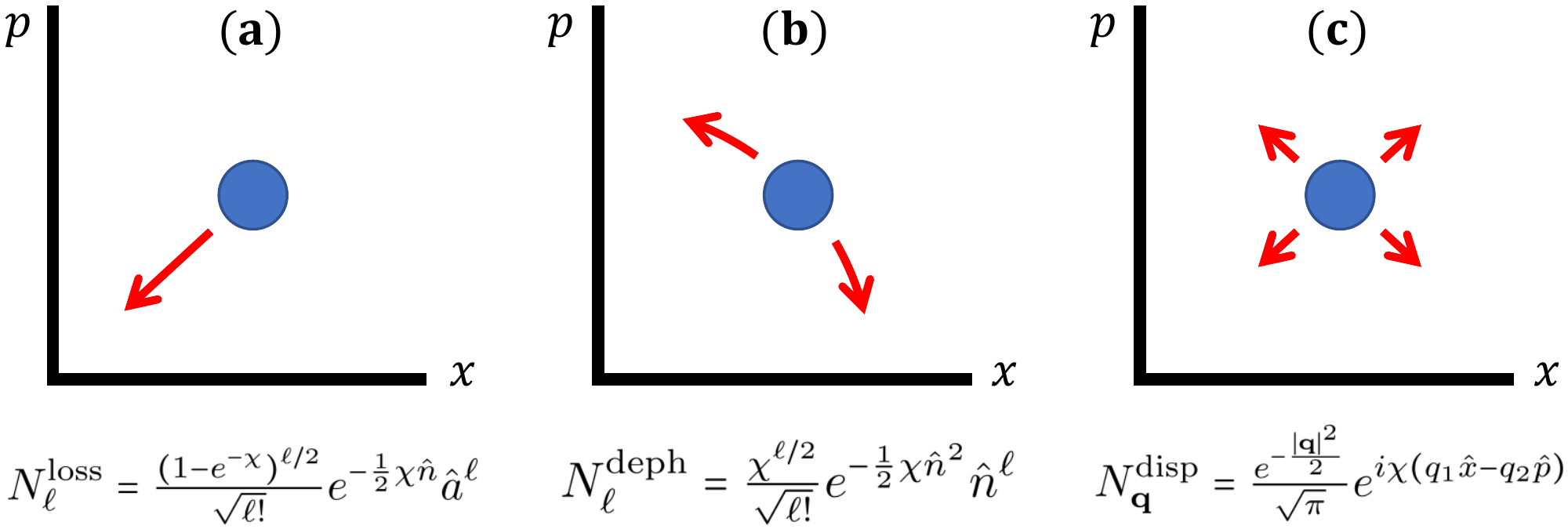}
\caption{Three types of common noise channels in bosonic systems --- \textbf{(a)} loss, \textbf{(b)} dephasing, and \textbf{(c)} displacement noise. The effect of each channel on a coherent state (blue circle) in \((x,p)\)-phase space is sketched using red arrows. Kraus operators are shown below, with \(\chi\) the noise strength, \(\ell\geq0\), and \(\mathbf{q}=(q_1,q_2)\).
}
\label{f:noise}
\end{figure}

Quantum information stored in a mode incurs one or more of the following main types of physical noise. Leakage of particles out of a mode (\textit{a.k.a.}\@ fiber attenuation) is represented by the \df{loss} channel (\textit{a.k.a.}\@ bosonic amplitude damping), dominant in electromagnetic modes~\cite{Girvin2021}. Cavity \df{dephasing}, whose errors are fluctuations of the oscillator phase and are functions of \(\nh\), is a prominent noise source in mechanical modes. We have already seen \df{displacement} noise (\textit{a.k.a.}\@ quantum additive Gaussian white noise or AGWN),
present in an optical fiber experiencing attenuation and amplification~\cite{Banaszek2020}.
Qualitatively, loss causes radial contraction, dephasing causes angular diffusion, and displacement noise causes uniform diffusion in phase space (see Fig.~\ref{f:noise}).

Any noise channel \(\N\) can be expressed in Kraus form, \(\N(\rho)=\sum_{\ell} N_\ell \rho N_\ell^\dagger\), with Kraus operators \(N_\ell\).
Loss Kraus operators, shown in Fig.~\ref{f:noise}(a), are proportional the \(\ell\)th power of the lowering operator \(\al\),
multiplied by a cooling operator (\ref{eq:cooling}) that drives the population of the system to the vacuum Fock state~\cite{Girvin2021}. Operators corresponding to \(\ell\)th-order dephasing noise, shown in Fig.~\ref{f:noise}(b), are proportional to the \(\ell\)th power of \(\nh\)~\cite{paircat}. Expanding exponentials of \(\nh\) in the loss and dephasing operators makes contact with loss and dephasing errors, \(\al^\ell\) and \(\nh^\ell\), respectively.

The strength of the noise is quantified by \(\chi\in[0,\infty)\).
Typically, \(\chi=\kappa t\), where \(t\) is how long the channel is ``turned on'', and \(\kappa\) is a noise rate. For optical fibers, \(\chi\) is the fiber's length divided by its attenuation length.
For displacement noise, \(\kappa t=\sigma\) is the displacement distribution's standard deviation, discussed in the context of GKP-stabilizer codes in Sec. ~\ref{sec:gkp-stab}.

Suppose we want to use a particular mode to store or communicate information; which code do we pick? The answer is, as always, a function of (A) the system's noise profile, (B) the code's error-correcting properties, and (C) the engineer's level of control.
Leaving (C) to the more knowledgeable, I summarize work on (A-B).

GKP stabilizer codes, designed with displacement noise in mind, are suitable for AGWN channels. Multimode GKP codes can be used to achieve the communication capacity of the quantum AGWN channel, up to a constant offset~\cite{Harrington2001,Sharma2018,Rosati2018,Noh2018}.

Fock-state codes, protecting against shifts in the occupation number as well as dephasing errors, should be a suitable counter to loss and dephasing. Indeed, large-\(\alpha\) cat codes and related binomial and number-phase codes substantially outperform GKP codes against pure dephasing noise~\cite{Joshi2021,Leviant2022} (see also Ref.~\cite{Li2016}). A similar performance comparison shows \textit{the opposite} is true for pure loss~\cite{codecomp}.
This is because loss Kraus operators, when expanded in terms of displacements, are mostly supported by small, and therefore correctable, position and momentum shifts~\cite[Eq. (7.12)]{codecomp}.
There is a more transparent effect on the channel level: since loss combined with \df{amplification} --- a channel proportional to the adjoint of loss --- yields displacement noise, amplification followed by GKP stabilizer-based recovery is a well-performing decoder.
In fact, GKP codes \textit{also} achieve the capacity of pure loss and related channels~\cite{Sharma2018,Rosati2018,Noh2018}.

Combinations of non-negligible loss and dephasing may prove too noisy for error correction to be useful: optimization favors
codewords supported on the first few Fock states~\cite{dephasing,Leviant2022}, close to the trivial Fock-state encoding \(\{|n=0\ket,|n=1\ket\}\). Similar results are seen when loss is combined with coherent errors due to the Kerr Hamiltonian \(\nh^2\)~\cite{codecomp}.

\section{Use cases of bosonic encodings}
\label{sec:advantage}

Why should we consider bosonic platforms (and their associated codes) over others?
For one thing, such platforms are sometimes the \textit{only} ones available for a given task.
For example, if one wants to communicate quantum information over long distances, one would find it very difficult to avoid dealing with either optical-fiber or free-space quantum links.
But besides such ubiquity,
bosonic platforms also yield unique advantages.
I conclude by highlighting several ongoing developments.

\subsection{Hardware efficiency}

Use of bosonic platforms for a particular task may be \df{hardware efficient}, which means, loosely speaking, that a bosonic platform requires fewer physical resources to perform a task than other platforms.

A small bosonic code can maintain the same degree of protection as a few-qubit code, but require fewer quantum states.
For example,
the ``0-2-4'' binomial code (\ref{eq:binomial}) is more hardware efficient than the four-qubit code~\cite[Table 4]{Girvin2021}.
Both codes protect against the same type of noise --- amplitude damping (qubit and bosonic, respectively) --- to first order in the damping parameter \(\chi\) (against the operators \(\{I,\sigma_-\)\} and \(\{I,\al\}\), respectively). However, the binomial code requires only \(5\) Fock states and \(3\) error spaces, compared to the four-qubit code's \(2^4=16\) states and \(2^3=8\) error spaces.

For a more drastic comparison, families of cat codes are more hardware efficient than families of repetition codes.
Both codes offer excellent protection against one type of noise: cat-code schemes suppress dephasing noise exponentially with \(\alpha^2\) [see Eq. (\ref{eq:cat-dephasing})], while \(n\)-qubit repetition-code schemes suppress bit-flip noise exponentially with \(n\). However, repetition codes require an exponentially increasing \(n\)-qubit space, while cat codes fit into a single mode for any \(\alpha\).
The hardware-efficient stability offered by cat codes has been linked to several favorable properties of phases of matter associated with the repetition-code Hamiltonian --- the classical 2D Ising model~\cite{Minganti2018,Lieu2020} --- and continues to be actively studied~\cite{Lieu2022,Gravina2022}.

\subsection{Boosting performance with analog decoding}
\label{sec:threshold}

A multi-qubit code can be concatenated with a single-mode bosonic code to yield a better performing multi-mode code, at the expense of the extra bosonic overhead. In the typical example of such a concatenation, each physical qubit of an outer \([[n,k,d]]\) qubit stabilizer code~\cite{gottesman_thesis,Calderbank1996}\zoo{qubit_stabilizer} is further encoded in an inner single-mode bosonic code.

To take advantage of the concatenation, the qubit code's decoder, which uses only \textit{digital} information from physical qubit measurements, can be modified to take in the \textit{soft} or \textit{analog} information~\cite{Pattison2021} of the often continuous syndromes of the bosonic code.
The resulting performance boost is evident even for small codes, e.g., a certain decoder for a three-mode repetition code concatenated with GKP qubits can accommodate some two-mode errors~\cite{Fukui2017a}.

An important feature of infinite families of qubit codes is the existence of a \df{threshold} --- the largest strength of a given physical noise channel for which the inaccuracy of correction \(\epsilon\) [see Eq. (\(\ref{eq:qec}\))] decreases exponentially with the distance \(d\) of the code family.
Threshold values enable engineers to predict the maximum noisiness that can be tolerated with error correction.
A multi-qubit \(\{[[n_i,k_i,d_i]]\}_{i\geq 1}\) code family admitting a threshold can be concatenated with a bosonic code, such as the two-component cat~\cite{Puri2019,Chamberland2020,Darmawan2021} or square-lattice GKP
\cite{Vuillot2018,Noh2019,Larsen2021,Hanggli2020,Noh2021} code, to increase the threshold. Such concatenation schemes are compatible not only with gate-based computation, but also with alternative measurement-based approaches tailored to optical platforms
\cite{Menicucci2013a,Fukui2017b,Bourassa2020,Tzitrin2021}.

\subsection{Continuous transversal gates}

A \df{transversal gate} acting on an \(n\)-qubit code is a gate that can be implemented as a tensor product of unitary operations, each acting on a subset of qubits whose size is independent of \(n\).
Such gates are particularly useful for fault tolerance because any single-qubit error occurring before a transversal gate can only spread to the few-qubit subset that interacts with that qubit via the gate.

The Eastin-Knill theorem~\cite{Eastin2008} states that a finite-dimensional multi-qubit code detecting few-qubit errors cannot have a continuous-parameter set of transversal gates. The argument is that any purported transversal gates would be exponentials of a sum of few-qubit terms corresponding to each qubit subset, and such terms would yield identity when projected into the codespace since the code is assumed to detect few-qubit noise.

Due to infinite dimensionality, bosonic modes offer one way to circumvent the Eastin-Knill theorem. For example, the four-mode code (\ref{eq:four-mode}) admits an infinite family of logical displacements \(\{e^{-iq(\pp_1+\pp_2)}\}_{q\in\R}\). More involved codes exist for oscillators, rotors, and other spaces~\cite{Faist2019}, allowing for arbitrarily large transversal-gate families. Such gates also hold for ``cooled'' normalizable versions of exact codewords, albeit the error correction becomes approximate (see Sec.~\ref{sec:bosonic-stab}).

\subsection{Noise-bias preservation}

The interplay of gates with noise is important for fault tolerance.
Generic gates often change the \textit{type} of noise upon permutation, e.g., converting bit to phase noise.
\df{Bias-preserving gates}~\cite{Aliferis2007} are gates that \textit{do not} change the type of noise upon such a permutation. Such gates are useful for maintaining fault tolerance in systems whose noise profile is biased toward either bit or phase noise.

It is impossible to obtain a \(\cnot\) gate using a Hamiltonian-based bias-preserving rotation for qubit codes~\cite{Aliferis2007}.
For any unitary \(e^{i s\hat{H}}\) generated by a two-qubit Hamiltonian \(\hat{H}\) such that \(e^{i s\hat{H}}|_{s=1}=\cnot\), there is an intermediate \(s\) at which the unitary converts phase-flip to bit-flip noise~\cite[Appx. A]{Guillaud2019}.

Due to infinite dimensionality, oscillator codes can circumvent this no-go theorem, implementing continuous Hamiltonian-based \(\cnot\) gates that maintain noise bias throughout.
A bias-preserving logical \(\cnot\) gate exists for the ``two-component'' cat code~\cite{Puri2019,Guillaud2019}. It is based on the logical Pauli gate \(e^{is\pi\nh}|_{s=1}\) that permutes the code's antipodal coherent-state codewords, \(|\alpha\ket\) and \(\left|-\alpha\right\rangle\), while maintaining the phase-space distance between them for all \(s\) (see notes~\cite{leshouches_shruti}).
The logical \(\cnot\) gate for GKP qubits, achieved by a Hamiltonian-based symplectic transformation, similarly preserves noise bias in either the position or momentum quadratures \cite[Eq.~(104)]{Gottesman2001}.

Due to the ability to preserve bias, cat codes concatenated with particular stabilizer codes yield substantial increases in performance for systems whose noise is substantially biased~\cite{cohenthesis,Guillaud2019,Chamberland2020,Darmawan2021}.
The bias in cat codes also makes for a good ancilla, yielding fault-tolerant syndrome measurement schemes~\cite{Puri2018}.

\subsection{GKP states as non-Gaussian resources}

Multi-qubit codes typically admit a set of ``easy'' gates called the Clifford-group gates, and a ``hard'' non-Clifford element is required to achieve universal computation.
The same relationship holds between GKP codes and \df{Gaussian operations} --- ``easy'' oscillator unitaries consisting of Fock-space rotations, displacements, and two other elements called beam splitters and squeezers~\cite{Cerf2007,serafinibook}.
Gaussian operations allow for only logical Clifford operations on encoded GKP qubits,
so a non-Gaussian resource is required for universal quantum computation.

In related fashion, almost all bosonic codewords are outside of the set of \df{Gaussian states}, which can be made by applying Gaussian operations on the vacuum Fock state or the thermal state \(e^{-\beta \nh}\). The important exception is analog stabilizer codewords [e.g., the four-mode codewords (\ref{eq:four-mode})], which are generated using only Gaussian elements acting on position states, which in turn can be thought of as infinitely squeezed vacuum states.

\df{Magic states} are non-Gaussian resource states that are used to perform non-Clifford GKP logical gates via gate teleportation.
Magic states are typically generated in a separate and often cumbersome distillation process requiring other non-Gaussian resources.
This is not the case for GKP codes, whose magic states can be distilled directly by applying GKP error correction to Gaussian initial states, circumventing the need for non-Gaussian elements other than GKP error correction~\cite{Baragiola2019}.

GKP-stabilizer codes can be thought of as analog stabilizer codes ``upgraded'' with a non-Gaussian resource, allowing for error correction that is not possible with purely Gaussian resources.
Analog stabilizer codewords can be initialized by applying a Gaussian circuit to tensor-product states consisting of the desired quantum information \(|\psi\ket\) in the first \(k\) physical mode and position states \(|x=0\ket\) in the remaining modes.
GKP-stabilizer codes can be initialized in the same way, but with the substitution of canonical GKP states (\ref{eq:gkpcomb}) for the initial position states.

The GKP resource is what allows GKP-stabilizer codes to protect an entire mode against Gaussian shifts, as there is a no-go theorem~\cite{Niset2008a,Vuillot2018} showing that analog stabilizer codes cannot protect against such shifts (see also~\cite{Giedke2002,Eisert2002}).
However, there are no such restrictions for analog stabilizer codes against \textit{non}-Gaussian noise~\cite{VanLoock2010}.
Moreover, the existence of a threshold for protecting logical modes against Gaussian displacement noise using GKP-stabilizer codes is an open question and at the very least requires ideal (i.e., non-normalizable) codewords~\cite{Hanggli2021}.

\subsection{Other curiosities}

Bosonic modes may be helpful in providing simple (i.e., commuting-projector) models for exotic \textit{topological} phases of quantum matter, such as fractional quantum Hall systems.
This task is shown to be impossible for systems consisting of finite-dimensional qudits~\cite{Kapustin2018}, and there has been some progress in developing models on planar rotors
\cite{DeMarco2021a,DeMarco2021}.

Bosonic modes may be useful for quantum simulation of, unsurprisingly, other bosonic modes.
Boson sampling~\cite{Aaronson2010}, a fundamentally difficult problem in terms of its computational complexity, is naturally baked into bosonic degrees of freedom.
Decoding of multimode GKP codes is interestingly tied to gauge theory~\cite{Vuillot2018}.
Recent work has shown how to extract parameters relevant to chemical processes involving molecular vibrational modes~\cite{Wang2019}, although work on quantum-inspired classical algorithms has fired back at the claim that a quantum device makes the extraction more efficient~\cite{Oh2022}.

Infinite dimensions also bear fruit for quantum control. Amplifying the time of evolution under the harmonic oscillator Hamiltonian is possible using a sequence of single-mode squeezing operations, but similar amplifications are not possible in finite dimensions~\cite{Arenz2018}.

\section{Summary and discussion}\label{sec:summary}

The primary application of bosonic codes is the storage and processing quantum information in phononic and photonic devices.
Due to their formally infinite-dimensional space, bosonic systems yield quantum protocols that are forbidden in the discrete-variable (DV) setting.
I review bosonic memories and highlight several such advantageous protocols.

The primary goal of error correction is to yield a probability of corruption of the encoded information (\(p_{\text{logical}}\)) that is smaller than a reference probability of unencoded information corruption (\(p_{\text{physical}}\)). Defining the \df{coding gain} of a given error correcting protocol as the ratio of the two probabilities, an error-correcting protocol is useful if it yields a gain larger than one.
A more ambitious goal in the quantum setting is made possible by the notion of an error-correcting threshold (\(p_{\text{threshold}}\)), below which it should be possible to construct error-correcting protocols with \textit{arbitrarily large} gain by scaling up the physical degrees of freedom (see Sec.~\ref{sec:threshold}). The equation below summarizes the three important regimes of the gain, which are independent of the quantitative formulations of the three probabilities:
\begin{equation}
\text{gain}=\frac{p_{\text{physical}}}{p_{\text{logical}}}\,\,\begin{cases}
~=1 & \text{break-even QEC}\\
~>1 & \text{beyond break-even}\\
~\to\infty & \text{below threshold,}~p_{\text{physical}} < p_{\text{threshold}}
\end{cases}~.
\end{equation}

Bosonic codes have initially outpaced DV codes in terms of approaching and surpassing unity gain. The first demonstration of break-even correction was achieved using cat codes several years ago~\cite{Ofek2016}, and the first substantial greater-that-unity gain has been achieved recently using GKP codes~\cite{SivakVolodymyr}. Qubit platforms are quickly catching up: a recent trapped-ion qubit implementation has made progress toward going beyond break-even for two-qubit gates~\cite{Ryan-Anderson2022}.

Few-qubit and few-mode bosonic codes are limited by a ceiling in the gain because the only (known) way of achieving arbitrarily large gain is to be below the threshold physical error rate of an infinite family of codes whose Hilbert-space dimension increases exponentially and which admits a threshold.
In accordance with that notion, large-scale industrial experiments instead focus on observing the proper scaling of the gain with system size~\cite{Chen2021a,Acharya2022}, betting that a higher-than-unity gain would eventually follow once such scaling is established.

The future of quantum error correction likely lies in the synthesis of both bosonic and DV error correction paradigms, resulting in multiple layers of protection tailored to specific devices and applications. Bosonic and DV strategies are not competing technologies, but two pieces of a puzzle that should be integrated to maximize progress.

\acknowledgments
These notes are not meant to review the large bosonic code literature, and I apologize for missing references that would have been relevant to the topics discussed here. On the other hand, the error-correction zoo
\pdfstartlink attr{/Border[0 0 0]} user{/Subtype /Link /A << /S /URI /URI (https://errorcorrectionzoo.org/) >>}{\includegraphics[height=10pt,keepaspectratio]{zoo_icon.png}}\pdfendlink\@ \textit{is} meant to be a comprehensive collection, so please notify me of any references that should be included on that site, or add them yourself on our Github page
\pdfstartlink attr{/Border[0 0 0]} user{/Subtype /Link /A << /S /URI /URI (https://github.com/errorcorrectionzoo) >>}{\includegraphics[height=10pt,keepaspectratio]{zoo_icon.png}}\pdfendlink.

These notes have benefitted greatly from co-teaching courses on classical and quantum error correction with M. Gullans and A. Barg in the Fall and Spring of the past academic year, respectively, using a draft of D. Gottesman's error-correction book. I gratefully acknowledge discussions with numerous colleagues, including R. Alexander, C. Arenz, B. Baragiola, A. Barg, R. Blume-Kohout, K. Brown, J. Conrad, M. H. Devoret, C. Eichler, P. Faist, T. Gerrits, S. M. Girvin, A. Grimm, D. Gottesman, J. Iosue, J. Iverson, H. Jeong, L. Jiang, K. Noh, H. Pfister, J. Preskill, S. Puri, B. Royer, R. J. Schoelkopf, V. Sivak, B. Terhal, and A. Wallraff.
I thank O. Albert and R. Kandratsenia for providing daycare support during the drafting of this manuscript.

I acknowledge support from an NSF QLCI (award No. OMA-2120757).
Contributions to this work by NIST, an agency of the US government, are not subject to US copyright. Any mention of commercial products does not indicate endorsement by NIST.

\bibliographystyle{varenna}
\bibliography{C:/Users/russi/Documents/library}

\begin{thebibliography}{100}
\expandafter\ifx\csname url\endcsname\relax\def\url#1{\texttt{#1}}\fi
\expandafter\ifx\csname urlprefix\endcsname\relax\def\urlprefix{URL }\fi

\bibitem{Shannon1948}
\NAME{Shannon C.~E.}, \IN{Bell Syst. Tech. J.}{27}{1948}{379}.
\newline\urlprefix\url{https://ieeexplore.ieee.org/document/6773024}

\bibitem{Anderson2002}
\NAME{Anderson J.~B. \atque Svensson A.}, \TITLE{{Coded Modulation Systems}},
  Information Technology: Transmission, Processing and Storage (Kluwer Academic
  Publishers, Boston) 2002.
\newline\urlprefix\url{http://link.springer.com/10.1007/b100498}

\bibitem{Lapidoth2017}
\NAME{Lapidoth A.}, \TITLE{{A Foundation in Digital Communication}} (Cambridge
  University Press) 2017.
\newline\urlprefix\url{https://www.cambridge.org/core/product/identifier/9781316822708/type/book}

\bibitem{Girvin2021}
\NAME{Girvin S.~M.}, \IN{e-print}{}{2021}{}.
\newline\urlprefix\url{http://arxiv.org/abs/2111.08894}

\bibitem{preskillnotes}
\NAME{Preskill J.}, \TITLE{{Lecture notes on Quantum Computation}} (online
  notes).
\newline\urlprefix\url{http://www.theory.caltech.edu/people/preskill/ph229/}

\bibitem{Nielsen2011}
\NAME{Nielsen M.~A. \atque Chuang I.~L.}, \TITLE{{Quantum Computation and
  Quantum Information: 10th Anniversary Edition}}, 10th Edition (Cambridge
  University Press, New York, NY, USA) 2011.

\bibitem{Terhal2015}
\NAME{Terhal B.~M.}, \IN{Rev. Mod. Phys.}{87}{2015}{307}.
\newline\urlprefix\url{http://link.aps.org/doi/10.1103/RevModPhys.87.307}

\bibitem{gottesmanbook}
\NAME{Gottesman D.}, \TITLE{{Surviving as a quantum computer in a classical
  world}} (unpublished).

\bibitem{Joshi2021}
\NAME{Joshi A., Noh K. \atque Gao Y.~Y.}, \IN{Quantum Sci.
  Technol.}{6}{2021}{033001}.
\newline\urlprefix\url{https://iopscience.iop.org/article/10.1088/2058-9565/abe989}

\bibitem{Cai2020}
\NAME{Cai W., Ma Y., Wang W., Zou C.-L. \atque Sun L.}, \IN{Fundam.
  Res.}{1}{2021}{50}.
\newline\urlprefix\url{http://arxiv.org/abs/2010.08699
  http://dx.doi.org/10.1016/j.fmre.2020.12.006
  https://linkinghub.elsevier.com/retrieve/pii/S2667325820300145}

\bibitem{Terhal2020}
\NAME{Terhal B.~M., Conrad J. \atque Vuillot C.}, \IN{Quantum Sci.
  Technol.}{5}{2020}{043001}.
\newline\urlprefix\url{http://arxiv.org/abs/2002.11008
  http://dx.doi.org/10.1088/2058-9565/ab98a5
  https://iopscience.iop.org/article/10.1088/2058-9565/ab98a5}

\bibitem{Noh2021a}
\NAME{Noh K.}, \IN{}{}{2021}{}.
\newline\urlprefix\url{http://arxiv.org/abs/2103.09445}

\bibitem{Guillaud2022}
\NAME{Guillaud J., Cohen J. \atque Mirrahimi M.}, \IN{}{}{2022}{}.
\newline\urlprefix\url{http://arxiv.org/abs/2203.03222}

\bibitem{Cerf2007}
\NAME{Cerf N.~J., Leuchs G. \atque Polzik E.~S.}, \TITLE{{Quantum Information
  with Continuous Variables of Atoms and Light}} (PUBLISHED BY IMPERIAL COLLEGE
  PRESS AND DISTRIBUTED BY WORLD SCIENTIFIC PUBLISHING CO.) 2007.
\newline\urlprefix\url{https://www.worldscientific.com/worldscibooks/10.1142/p489}

\bibitem{serafinibook}
\NAME{Serafini A.}, \TITLE{{Quantum Continuous Variables: A Primer of
  Theoretical Methods}} (CRC Press, Boca Raton FL) 2017.
\newline\urlprefix\url{https://www.crcpress.com/Quantum-Continuous-Variables-A-Primer-of-Theoretical-Methods/Serafini/p/book/9781482246346}

\bibitem{Gottesman2009}
\NAME{Gottesman D.}, \IN{}{}{2009}{}.
\newline\urlprefix\url{http://arxiv.org/abs/0904.2557}

\bibitem{Knill1997}
\NAME{Knill E. \atque Laflamme R.}, \IN{Phys. Rev. A}{55}{1997}{900}.
\newline\urlprefix\url{http://journals.aps.org/pra/abstract/10.1103/PhysRevA.55.900}

\bibitem{gottesman_thesis}
\NAME{Gottesman D.}, \TITLE{{Stabilizer codes and quantum error correction}},
  Ph.D. thesis, California Institute of Technology (1997).
\newline\urlprefix\url{https://arxiv.org/abs/quant-ph/9705052}

\bibitem{Calderbank1996}
\NAME{Calderbank A.~R., Rains E.~M., Shor P.~W. \atque Sloane N. J.~A.},
  \IN{Phys. Rev. Lett.}{78}{1996}{405}.
\newline\urlprefix\url{http://arxiv.org/abs/quant-ph/9605005
  http://dx.doi.org/10.1103/PhysRevLett.78.405}

\bibitem{Vaidman1996}
\NAME{Vaidman L., Goldenberg L. \atque Wiesner S.}, \IN{Phys. Rev.
  A}{54}{1996}{R1745}.
\newline\urlprefix\url{https://link.aps.org/doi/10.1103/PhysRevA.54.R1745}

\bibitem{scully_book}
\NAME{Scully M.~O. \atque Zubairy M.~S.}, \TITLE{{Quantum Optics}} (Cambridge
  University Press, Cambridge) 1997.
\newline\urlprefix\url{https://www.amazon.com/Quantum-Optics-Marlan-Scully/dp/0521435951}

\bibitem{Artiles2005}
\NAME{Artiles L.~M., Gill R.~D. \atque Guta M.~I.}, \IN{J. R. Stat. Soc. Ser. B
  (Statistical Methodol.}{67}{2005}{109}.
\newline\urlprefix\url{http://arxiv.org/abs/math/0405595
  https://onlinelibrary.wiley.com/doi/10.1111/j.1467-9868.2005.00491.x}

\bibitem{Lvovsky2008}
\NAME{Lvovsky A.~I. \atque Raymer M.~G.}, \IN{Rev. Mod. Phys.}{}{2008}{1}.

\bibitem{Holevo2011}
\NAME{Holevo A.}, \TITLE{{Probabilistic and Statistical Aspects of Quantum
  Theory}} (Edizioni della Normale, Pisa) 2011.
\newline\urlprefix\url{http://link.springer.com/10.1007/978-88-7642-378-9}

\bibitem{Gieres2000}
\NAME{Gieres F.}, \IN{Reports Prog. Phys.}{63}{2000}{1893}.
\newline\urlprefix\url{http://arxiv.org/abs/quant-ph/9907069
  http://dx.doi.org/10.1088/0034-4885/63/12/201
  https://iopscience.iop.org/article/10.1088/0034-4885/63/12/201}

\bibitem{Madrid2005}
\NAME{de~la Madrid R.}, \IN{Eur. J. Phys.}{26}{2005}{287}.
\newline\urlprefix\url{http://arxiv.org/abs/quant-ph/0502053
  http://dx.doi.org/10.1088/0143-0807/26/2/008
  https://iopscience.iop.org/article/10.1088/0143-0807/26/2/008}

\bibitem{Iosue22}
\NAME{Iosue J.~T., Sharma K., Gullans M.~J. \atque Albert V.~V.},
  \IN{e-print}{}{2022}{}.

\bibitem{nielsen_chuang}
\NAME{Nielsen M.~A. \atque Chuang I.~L.}, \TITLE{{Quantum Computation and
  Quantum Information}} (Cambridge University Press, Cambridge) 2011.
\newline\urlprefix\url{http://www.amazon.com/Quantum-Computation-Information-Anniversary-Edition/dp/1107002176}

\bibitem{Barnes2004}
\NAME{Barnes R.~L.}, \IN{e-print}{}{2004}{}.
\newline\urlprefix\url{http://arxiv.org/abs/quant-ph/0405064}

\bibitem{Cahill1969}
\NAME{Cahill K.~E. \atque Glauber R.~J.}, \IN{Phys. Rev.}{177}{1969}{1857}.
\newline\urlprefix\url{https://link.aps.org/doi/10.1103/PhysRev.177.1857}

\bibitem{Lloyd1998}
\NAME{Lloyd S. \atque Slotine J.-J.~E.}, \IN{Phys. Rev. Lett.}{80}{1998}{4088}.
\newline\urlprefix\url{https://link.aps.org/doi/10.1103/PhysRevLett.80.4088}

\bibitem{Braunstein1998}
\NAME{Braunstein S.~L.}, \IN{Phys. Rev. Lett.}{80}{1998}{4084}.
\newline\urlprefix\url{https://link.aps.org/doi/10.1103/PhysRevLett.80.4084}

\bibitem{Peres1985}
\NAME{Peres A.}, \IN{Phys. Rev. A}{32}{1985}{3266}.
\newline\urlprefix\url{https://link.aps.org/doi/10.1103/PhysRevA.32.3266}

\bibitem{Aoki2009}
\NAME{Aoki T., Takahashi G., Kajiya T., Yoshikawa J.-i., Braunstein S.~L., van
  Loock P. \atque Furusawa A.}, \IN{Nat. Phys.}{5}{2009}{541}.
\newline\urlprefix\url{http://www.nature.com/articles/nphys1309}

\bibitem{Zhang2008}
\NAME{Zhang J., Xie C., Peng K. \atque van Loock P.}, \IN{Phys. Rev.
  A}{78}{2008}{052121}.
\newline\urlprefix\url{http://link.aps.org/doi/10.1103/PhysRevA.78.052121}

\bibitem{Menicucci2006}
\NAME{Menicucci N.~C., van Loock P., Gu M., Weedbrook C., Ralph T.~C. \atque
  Nielsen M.~A.}, \IN{Phys. Rev. Lett.}{97}{2006}{110501}.
\newline\urlprefix\url{https://link.aps.org/doi/10.1103/PhysRevLett.97.110501}

\bibitem{Gu2009a}
\NAME{Gu M., Weedbrook C., Menicucci N.~C., Ralph T.~C. \atque van Loock P.},
  \IN{Phys. Rev. A}{79}{2009}{062318}.
\newline\urlprefix\url{http://arxiv.org/abs/0903.3233
  http://dx.doi.org/10.1103/PhysRevA.79.062318
  https://link.aps.org/doi/10.1103/PhysRevA.79.062318}

\bibitem{Gottesman2001}
\NAME{Gottesman D., {Yu. Kitaev} A. \atque Preskill J.}, \IN{Phys. Rev.
  A}{64}{2001}{012310}.
\newline\urlprefix\url{http://link.aps.org/doi/10.1103/PhysRevA.64.012310}

\bibitem{Harrington2001}
\NAME{Harrington J. \atque Preskill J.}, \IN{Phys. Rev. A}{64}{2001}{062301}.
\newline\urlprefix\url{http://link.aps.org/doi/10.1103/PhysRevA.64.062301}

\bibitem{Noh2019a}
\NAME{Noh K., Girvin S.~M. \atque Jiang L.}, \IN{Phys. Rev.
  Lett.}{125}{2020}{080503}.
\newline\urlprefix\url{http://arxiv.org/abs/1903.12615
  https://link.aps.org/doi/10.1103/PhysRevLett.125.080503}

\bibitem{Xu2022}
\NAME{Xu Y., Wang Y., Kuo E.-J. \atque Albert V.~V.}, \IN{e-print}{}{2022}{}.
\newline\urlprefix\url{http://arxiv.org/abs/2209.04573}

\bibitem{Menicucci2014}
\NAME{Menicucci N.~C.}, \IN{Phys. Rev. Lett.}{112}{2014}{120504}.
\newline\urlprefix\url{https://link.aps.org/doi/10.1103/PhysRevLett.112.120504}

\bibitem{Terhal2016}
\NAME{Terhal B.~M. \atque Weigand D.}, \IN{Phys. Rev. A}{93}{2016}{012315}.
\newline\urlprefix\url{http://link.aps.org/doi/10.1103/PhysRevA.93.012315}

\bibitem{CampagneIbarcq2019}
\NAME{Campagne-Ibarcq P., Eickbusch A., Touzard S., Zalys-Geller E., Frattini
  N.~E., Sivak V.~V., Reinhold P., Puri S., Shankar S., Schoelkopf R.~J.,
  Frunzio L., Mirrahimi M. \atque Devoret M.~H.}, \IN{Nature}{584}{2020}{368}.
\newline\urlprefix\url{http://arxiv.org/abs/1907.12487
  https://www.nature.com/articles/s41586-020-2603-3}

\bibitem{DeNeeve2020}
\NAME{de~Neeve B., Nguyen T.-L., Behrle T. \atque Home J.~P.}, \IN{Nat.
  Phys.}{18}{2022}{296}.
\newline\urlprefix\url{http://arxiv.org/abs/2010.09681
  https://www.nature.com/articles/s41567-021-01487-7}

\bibitem{Royer2020}
\NAME{Royer B., Singh S. \atque Girvin S.~M.}, \IN{Phys. Rev.
  Lett.}{125}{2020}{260509}.
\newline\urlprefix\url{http://arxiv.org/abs/2009.07941
  http://dx.doi.org/10.1103/PhysRevLett.125.260509
  https://link.aps.org/doi/10.1103/PhysRevLett.125.260509}

\bibitem{Grimsmo2019}
\NAME{Grimsmo A.~L., Combes J. \atque Baragiola B.~Q.}, \IN{Phys. Rev.
  X}{10}{2020}{011058}.
\newline\urlprefix\url{http://arxiv.org/abs/1901.08071
  http://dx.doi.org/10.1103/PhysRevX.10.011058
  https://link.aps.org/doi/10.1103/PhysRevX.10.011058}

\bibitem{Albert2017}
\NAME{Albert V.~V., Pascazio S. \atque Devoret M.~H.}, \IN{J. Phys. A Math.
  Theor.}{50}{2017}{504002}.
\newline\urlprefix\url{http://stacks.iop.org/1751-8121/50/i=50/a=504002?key=crossref.d727101154c0cf4004bad5cf9a4ef386}

\bibitem{mol}
\NAME{Albert V.~V., Covey J.~P. \atque Preskill J.}, \IN{Phys. Rev.
  X}{10}{2020}{031050}.
\newline\urlprefix\url{http://arxiv.org/abs/1911.00099
  http://dx.doi.org/10.1103/PhysRevX.10.031050
  https://link.aps.org/doi/10.1103/PhysRevX.10.031050}

\bibitem{Susskind1964}
\NAME{Susskind L. \atque Glogower J.}, \IN{Phys. Phys. Fiz.}{1}{1964}{49}.
\newline\urlprefix\url{https://link.aps.org/doi/10.1103/PhysicsPhysiqueFizika.1.49}

\bibitem{Bergou1991}
\NAME{Bergou J. \atque Englert B.-G.}, \IN{Ann. Phys. (N. Y).}{209}{1991}{479}.
\newline\urlprefix\url{https://linkinghub.elsevier.com/retrieve/pii/0003491691900379}

\bibitem{dephasing}
\NAME{Combes J., Albert V.~V., Noh K., Woods M.~P., Grimsmo A.~L. \atque
  Baragiola B.~Q.}, \IN{(in Prep.}{}{2020}{}.

\bibitem{Ofek2016}
\NAME{Ofek N., Petrenko A., Heeres R., Reinhold P., Leghtas Z., Vlastakis B.,
  Liu Y., Frunzio L., Girvin S.~M., Jiang L., Mirrahimi M., Devoret M.~H.
  \atque Schoelkopf R.~J.}, \IN{Nature}{536}{2016}{441}.
\newline\urlprefix\url{http://www.nature.com/doifinder/10.1038/nature18949}

\bibitem{Rosenblum2018}
\NAME{Rosenblum S., Reinhold P., Mirrahimi M., Jiang L., Frunzio L. \atque
  Schoelkopf R.~J.}, \IN{Science (80-. ).}{361}{2018}{266}.
\newline\urlprefix\url{http://www.sciencemag.org/lookup/doi/10.1126/science.aat3996}

\bibitem{Helstrom1969}
\NAME{Helstrom C.~W.}, \IN{J. Stat. Phys.}{1}{1969}{231}.
\newline\urlprefix\url{http://link.springer.com/10.1007/BF01007479}

\bibitem{Leghtas2013b}
\NAME{Leghtas Z., Kirchmair G., Vlastakis B., Schoelkopf R.~J., Devoret M.~H.
  \atque Mirrahimi M.}, \IN{Phys. Rev. Lett.}{111}{2013}{120501}.
\newline\urlprefix\url{http://link.aps.org/doi/10.1103/PhysRevLett.111.120501}

\bibitem{codecomp}
\NAME{Albert V.~V., Noh K., Duivenvoorden K., Young D.~J., Brierley R.~T.,
  Reinhold P., Vuillot C., Li L., Shen C., Girvin S.~M., Terhal B.~M. \atque
  Jiang L.}, \IN{Phys. Rev. A}{97}{2018}{032346}.
\newline\urlprefix\url{https://link.aps.org/doi/10.1103/PhysRevA.97.032346}

\bibitem{Knill2003}
\NAME{Knill E.}, \IN{e-print}{}{2003}{}.
\newline\urlprefix\url{http://arxiv.org/abs/quant-ph/0312190}

\bibitem{Knill2004}
\NAME{Knill E.}, \IN{Nature}{434}{2005}{39}.
\newline\urlprefix\url{http://arxiv.org/abs/quant-ph/0410199
  http://dx.doi.org/10.1038/nature03350
  http://www.nature.com/articles/nature03350}

\bibitem{Wolinsky1988}
\NAME{Wolinsky M. \atque Carmichael H.~J.}, \IN{Phys. Rev.
  Lett.}{60}{1988}{1836}.
\newline\urlprefix\url{http://link.aps.org/doi/10.1103/PhysRevLett.60.1836}

\bibitem{Krippner1994}
\NAME{Krippner L., Munro W. \atque Reid M.}, \IN{Phys. Rev. A}{50}{1994}{4330}.
\newline\urlprefix\url{http://link.aps.org/doi/10.1103/PhysRevA.50.4330}

\bibitem{Hach1994}
\NAME{{Hach III} E. \atque Gerry C.}, \IN{Phys. Rev. A}{49}{1994}{490}.
\newline\urlprefix\url{http://link.aps.org/doi/10.1103/PhysRevA.49.490}

\bibitem{Gilles1994}
\NAME{Gilles L., Garraway B.~M. \atque Knight P.~L.}, \IN{Phys. Rev.
  A}{49}{1994}{2785}.
\newline\urlprefix\url{http://link.aps.org/doi/10.1103/PhysRevA.49.2785}

\bibitem{Paz1998}
\NAME{Paz J.~P. \atque Zurek W.~H.}, \IN{Proc. R. Soc. London. Ser. A Math.
  Phys. Eng. Sci.}{454}{1998}{355}.
\newline\urlprefix\url{http://rspa.royalsocietypublishing.org/content/454/1969/355
  https://royalsocietypublishing.org/doi/10.1098/rspa.1998.0165}

\bibitem{Cochrane1999}
\NAME{Cochrane P.~T., Milburn G.~J. \atque Munro W.~J.}, \IN{Phys. Rev.
  A}{59}{1999}{2631}.
\newline\urlprefix\url{http://link.aps.org/doi/10.1103/PhysRevA.59.2631}

\bibitem{Jeong2001}
\NAME{Jeong H. \atque Kim M.~S.}, \IN{Phys. Rev. A}{65}{2002}{042305}.
\newline\urlprefix\url{http://arxiv.org/abs/quant-ph/0109077
  http://dx.doi.org/10.1103/PhysRevA.65.042305
  https://link.aps.org/doi/10.1103/PhysRevA.65.042305}

\bibitem{Ralph2003}
\NAME{Ralph T.~C., Gilchrist A., Milburn G.~J., Munro W.~J. \atque Glancy S.},
  \IN{Phys. Rev. A}{68}{2003}{042319}.
\newline\urlprefix\url{http://link.aps.org/doi/10.1103/PhysRevA.68.042319}

\bibitem{Lund2007}
\NAME{Lund A.~P., Ralph T.~C. \atque Haselgrove H.~L.}, \IN{Phys. Rev.
  Lett.}{100}{2008}{030503}.
\newline\urlprefix\url{http://arxiv.org/abs/0707.0327
  http://dx.doi.org/10.1103/PhysRevLett.100.030503
  https://link.aps.org/doi/10.1103/PhysRevLett.100.030503}

\bibitem{Leghtas2014}
\NAME{Leghtas Z., Touzard S., Pop I.~M., Kou A., Vlastakis B., Petrenko A.,
  Sliwa K.~M., Narla A., Shankar S., Hatridge M.~J., Reagor M., Frunzio L.,
  Schoelkopf R.~J., Mirrahimi M. \atque Devoret M.~H.}, \IN{Science (80-.
  ).}{347}{2015}{}.
\newline\urlprefix\url{http://www.sciencemag.org/cgi/doi/10.1126/science.aaa2085}

\bibitem{S.Touzard}
\NAME{Touzard S., Grimm A., Leghtas Z., Mundhada S., Reinhold P., Axline C.,
  Reagor M., Chou K., Blumoff J., Sliwa K., Shankar S., Frunzio L., Schoelkopf
  R.~J., Mirrahimi M. \atque Devoret M.}, \IN{Phys. Rev. X}{8}{2018}{021005}.
\newline\urlprefix\url{https://link.aps.org/doi/10.1103/PhysRevX.8.021005}

\bibitem{Lescanne2019}
\NAME{Lescanne R., Villiers M., Peronnin T., Sarlette A., Delbecq M., Huard B.,
  Kontos T., Mirrahimi M. \atque Leghtas Z.}, \IN{Nat. Phys.}{16}{2020}{509}.
\newline\urlprefix\url{http://arxiv.org/abs/1907.11729
  http://dx.doi.org/10.1038/s41567-020-0824-x
  http://www.nature.com/articles/s41567-020-0824-x}

\bibitem{Puri2017}
\NAME{Puri S., Boutin S. \atque Blais A.}, \IN{npj Quantum Inf.}{3}{2017}{18}.
\newline\urlprefix\url{http://www.nature.com/articles/s41534-017-0019-1}

\bibitem{Grimm2019}
\NAME{Grimm A., Frattini N.~E., Puri S., Mundhada S.~O., Touzard S., Mirrahimi
  M., Girvin S.~M., Shankar S. \atque Devoret M.~H.},
  \IN{Nature}{584}{2020}{205}.
\newline\urlprefix\url{http://arxiv.org/abs/1907.12131
  http://dx.doi.org/10.1038/s41586-020-2587-z
  https://www.nature.com/articles/s41586-020-2587-z}

\bibitem{bin}
\NAME{Michael M.~H., Silveri M., Brierley R.~T., Albert V.~V., Salmilehto J.,
  Jiang L. \atque Girvin S.~M.}, \IN{Phys. Rev. X}{6}{2016}{031006}.
\newline\urlprefix\url{http://link.aps.org/doi/10.1103/PhysRevX.6.031006}

\bibitem{Layden2019}
\NAME{Layden D., Zhou S., Cappellaro P. \atque Jiang L.}, \IN{Phys. Rev.
  Lett.}{122}{2019}{040502}.
\newline\urlprefix\url{http://arxiv.org/abs/1811.01450
  http://dx.doi.org/10.1103/PhysRevLett.122.040502
  https://link.aps.org/doi/10.1103/PhysRevLett.122.040502}

\bibitem{Knill2001}
\NAME{Knill E., Laflamme R. \atque Milburn G.~J.}, \IN{Nature}{409}{2001}{46}.
\newline\urlprefix\url{http://www.nature.com/doifinder/10.1038/35051009}

\bibitem{Chuang1995}
\NAME{Chuang I.~L. \atque Yamamoto Y.}, \IN{Phys. Rev. A}{52}{1995}{3489}.
\newline\urlprefix\url{https://link.aps.org/doi/10.1103/PhysRevA.52.3489}

\bibitem{Chuang1997}
\NAME{Chuang I.~L., Leung D.~W. \atque Yamamoto Y.}, \IN{Phys. Rev.
  A}{56}{1997}{1114}.
\newline\urlprefix\url{http://link.aps.org/doi/10.1103/PhysRevA.56.1114}

\bibitem{Wasilewski2007}
\NAME{Wasilewski W. \atque Banaszek K.}, \IN{Phys. Rev. A}{75}{2007}{042316}.
\newline\urlprefix\url{https://link.aps.org/doi/10.1103/PhysRevA.75.042316}

\bibitem{Ouyang2018}
\NAME{Ouyang Y. \atque Chao R.}, \IN{IEEE Trans. Inf. Theory}{66}{2020}{2921}.
\newline\urlprefix\url{http://arxiv.org/abs/1809.09801
  http://dx.doi.org/10.1109/TIT.2019.2956142
  https://ieeexplore.ieee.org/document/8915836/}

\bibitem{Banaszek2020}
\NAME{Banaszek K., Kunz L., Jachura M. \atque Jarzyna M.}, \IN{J. Light.
  Technol.}{38}{2020}{2741}.
\newline\urlprefix\url{https://ieeexplore.ieee.org/document/8998224/}

\bibitem{paircat}
\NAME{Albert V.~V., Mundhada S.~O., Grimm A., Touzard S., Devoret M.~H. \atque
  Jiang L.}, \IN{Quantum Sci. Technol.}{4}{2019}{035007}.
\newline\urlprefix\url{http://dx.doi.org/10.1088/2058-9565/ab1e69}

\bibitem{Sharma2018}
\NAME{Sharma K., Wilde M.~M., Adhikari S. \atque Takeoka M.}, \IN{New J.
  Phys.}{20}{2018}{063025}.

\bibitem{Rosati2018}
\NAME{Rosati M., Mari A. \atque Giovannetti V.}, \IN{Nat.
  Commun.}{9}{2018}{4339}.
\newline\urlprefix\url{http://arxiv.org/abs/1801.04731
  http://dx.doi.org/10.1038/s41467-018-06848-0
  http://www.nature.com/articles/s41467-018-06848-0}

\bibitem{Noh2018}
\NAME{Noh K., Albert V.~V. \atque Jiang L.}, \IN{IEEE Trans. Inf.
  Theory}{65}{2019}{2563}.
\newline\urlprefix\url{https://ieeexplore.ieee.org/document/8482307/}

\bibitem{Leviant2022}
\NAME{Leviant P., Xu Q., Jiang L. \atque Rosenblum S.}, \IN{e-print}{}{2022}{}.
\newline\urlprefix\url{http://arxiv.org/abs/2205.00341}

\bibitem{Li2016}
\NAME{Li L., Zou C.-l., Albert V.~V., Muralidharan S., Girvin S.~M. \atque
  Jiang L.}, \IN{Phys. Rev. Lett.}{119}{2017}{030502}.
\newline\urlprefix\url{http://link.aps.org/doi/10.1103/PhysRevLett.119.030502}

\bibitem{Minganti2018}
\NAME{Minganti F., Biella A., Bartolo N. \atque Ciuti C.}, \IN{}{}{2018}{}.
\newline\urlprefix\url{http://arxiv.org/abs/1804.11293
  http://dx.doi.org/10.1103/PhysRevA.98.042118}

\bibitem{Lieu2020}
\NAME{Lieu S., Belyansky R., Young J.~T., Lundgren R., Albert V.~V. \atque
  Gorshkov A.~V.}, \IN{Phys. Rev. Lett.}{125}{2020}{240405}.
\newline\urlprefix\url{http://arxiv.org/abs/2008.02816
  http://dx.doi.org/10.1103/PhysRevLett.125.240405
  https://link.aps.org/doi/10.1103/PhysRevLett.125.240405}

\bibitem{Lieu2022}
\NAME{Lieu S., Liu Y.-J. \atque Gorshkov A.~V.}, \IN{e-print}{}{2022}{}.
\newline\urlprefix\url{http://arxiv.org/abs/2205.09767}

\bibitem{Gravina2022}
\NAME{Gravina L., Minganti F. \atque Savona V.}, \IN{e-print}{}{2022}{}.
\newline\urlprefix\url{http://arxiv.org/abs/2208.04928}

\bibitem{Pattison2021}
\NAME{Pattison C.~A., Beverland M.~E., da~Silva M.~P. \atque Delfosse N.},
  \IN{e-print}{}{2021}{}.
\newline\urlprefix\url{http://arxiv.org/abs/2107.13589}

\bibitem{Fukui2017a}
\NAME{Fukui K., Tomita A. \atque Okamoto A.}, \IN{Phys. Rev.
  Lett.}{119}{2017}{180507}.
\newline\urlprefix\url{http://arxiv.org/abs/1706.03011
  http://dx.doi.org/10.1103/PhysRevLett.119.180507
  https://link.aps.org/doi/10.1103/PhysRevLett.119.180507}

\bibitem{Puri2019}
\NAME{Puri S., St-Jean L., Gross J.~A., Grimm A., Frattini N.~E., Iyer P.~S.,
  Krishna A., Touzard S., Jiang L., Blais A., Flammia S.~T. \atque Girvin
  S.~M.}, \IN{Sci. Adv.}{6}{2020}{}.
\newline\urlprefix\url{http://arxiv.org/abs/1905.00450
  http://dx.doi.org/10.1126/sciadv.aay5901
  https://www.science.org/doi/10.1126/sciadv.aay5901}

\bibitem{Chamberland2020}
\NAME{Chamberland C., Noh K., Arrangoiz-Arriola P., Campbell E.~T., Hann C.~T.,
  Iverson J., Putterman H., Bohdanowicz T.~C., Flammia S.~T., Keller A., Refael
  G., Preskill J., Jiang L., Safavi-Naeini A.~H., Painter O. \atque
  Brand{\~{a}}o F.~G.}, \IN{PRX Quantum}{3}{2022}{010329}.
\newline\urlprefix\url{http://arxiv.org/abs/2012.04108
  http://dx.doi.org/10.1103/PRXQuantum.3.010329
  https://link.aps.org/doi/10.1103/PRXQuantum.3.010329}

\bibitem{Darmawan2021}
\NAME{Darmawan A.~S., Brown B.~J., Grimsmo A.~L., Tuckett D.~K. \atque Puri
  S.}, \IN{}{}{2021}{}.
\newline\urlprefix\url{http://arxiv.org/abs/2104.09539
  http://dx.doi.org/10.1103/PRXQuantum.2.030345}

\bibitem{Vuillot2018}
\NAME{Vuillot C., Asasi H., Wang Y., Pryadko L.~P. \atque Terhal B.~M.},
  \IN{Phys. Rev. A}{99}{2019}{032344}.
\newline\urlprefix\url{http://arxiv.org/abs/1810.00047
  http://dx.doi.org/10.1103/PhysRevA.99.032344
  https://link.aps.org/doi/10.1103/PhysRevA.99.032344}

\bibitem{Noh2019}
\NAME{Noh K. \atque Chamberland C.}, \IN{Phys. Rev. A}{101}{2020}{012316}.
\newline\urlprefix\url{http://arxiv.org/abs/1908.03579
  http://dx.doi.org/10.1103/PhysRevA.101.012316
  https://link.aps.org/doi/10.1103/PhysRevA.101.012316}

\bibitem{Larsen2021}
\NAME{Larsen M.~V., Chamberland C., Noh K., Neergaard-Nielsen J.~S. \atque
  Andersen U.~L.}, \IN{PRX Quantum}{2}{2021}{030325}.
\newline\urlprefix\url{http://arxiv.org/abs/2101.03014
  http://dx.doi.org/10.1103/PRXQuantum.2.030325
  https://link.aps.org/doi/10.1103/PRXQuantum.2.030325}

\bibitem{Hanggli2020}
\NAME{H{\"{a}}nggli L., Heinze M. \atque K{\"{o}}nig R.}, \IN{Phys. Rev.
  A}{102}{2020}{052408}.
\newline\urlprefix\url{https://link.aps.org/doi/10.1103/PhysRevA.102.052408}

\bibitem{Noh2021}
\NAME{Noh K., Chamberland C. \atque Brand{\~{a}}o F.~G.}, \IN{PRX
  Quantum}{3}{2022}{010315}.
\newline\urlprefix\url{http://arxiv.org/abs/2103.06994
  http://dx.doi.org/10.1103/PRXQuantum.3.010315
  https://link.aps.org/doi/10.1103/PRXQuantum.3.010315}

\bibitem{Menicucci2013a}
\NAME{Menicucci N.~C.}, \IN{Phys. Rev. Lett.}{112}{2014}{120504}.
\newline\urlprefix\url{http://arxiv.org/abs/1310.7596
  http://dx.doi.org/10.1103/PhysRevLett.112.120504
  https://link.aps.org/doi/10.1103/PhysRevLett.112.120504}

\bibitem{Fukui2017b}
\NAME{Fukui K., Tomita A., Okamoto A. \atque Fujii K.}, \IN{Phys. Rev.
  X}{8}{2018}{021054}.
\newline\urlprefix\url{http://arxiv.org/abs/1712.00294
  http://dx.doi.org/10.1103/PhysRevX.8.021054
  https://link.aps.org/doi/10.1103/PhysRevX.8.021054}

\bibitem{Bourassa2020}
\NAME{Bourassa J.~E., Alexander R.~N., Vasmer M., Patil A., Tzitrin I.,
  Matsuura T., Su D., Baragiola B.~Q., Guha S., Dauphinais G., Sabapathy K.~K.,
  Menicucci N.~C. \atque Dhand I.}, \IN{Quantum}{5}{2021}{392}.
\newline\urlprefix\url{http://arxiv.org/abs/2010.02905
  http://dx.doi.org/10.22331/q-2021-02-04-392
  https://quantum-journal.org/papers/q-2021-02-04-392/}

\bibitem{Tzitrin2021}
\NAME{Tzitrin I., Matsuura T., Alexander R.~N., Dauphinais G., Bourassa J.~E.,
  Sabapathy K.~K., Menicucci N.~C. \atque Dhand I.}, \IN{PRX
  Quantum}{2}{2021}{040353}.
\newline\urlprefix\url{http://arxiv.org/abs/2104.03241
  http://dx.doi.org/10.1103/PRXQuantum.2.040353
  https://link.aps.org/doi/10.1103/PRXQuantum.2.040353}

\bibitem{Eastin2008}
\NAME{Eastin B. \atque Knill E.}, \IN{Phys. Rev. Lett.}{102}{2009}{110502}.
\newline\urlprefix\url{http://arxiv.org/abs/0811.4262
  http://dx.doi.org/10.1103/PhysRevLett.102.110502
  https://link.aps.org/doi/10.1103/PhysRevLett.102.110502}

\bibitem{Faist2019}
\NAME{Faist P., Nezami S., Albert V.~V., Salton G., Pastawski F., Hayden P.
  \atque Preskill J.}, \IN{Phys. Rev. X}{10}{2020}{041018}.
\newline\urlprefix\url{http://arxiv.org/abs/1902.07714
  https://link.aps.org/doi/10.1103/PhysRevX.10.041018}

\bibitem{Aliferis2007}
\NAME{Aliferis P. \atque Preskill J.}, \IN{Phys. Rev. A}{78}{2008}{052331}.
\newline\urlprefix\url{http://arxiv.org/abs/0710.1301
  http://dx.doi.org/10.1103/PhysRevA.78.052331
  https://link.aps.org/doi/10.1103/PhysRevA.78.052331}

\bibitem{Guillaud2019}
\NAME{Guillaud J. \atque Mirrahimi M.}, \IN{Phys. Rev. X}{9}{2019}{041053}.
\newline\urlprefix\url{http://arxiv.org/abs/1904.09474
  http://dx.doi.org/10.1103/PhysRevX.9.041053
  https://link.aps.org/doi/10.1103/PhysRevX.9.041053}

\bibitem{leshouches_shruti}
\NAME{Puri S.}, \TITLE{{QEC when the noise is biased}} (online notes) 2019.
\newline\urlprefix\url{https://physinfo.fr/houches2019/program.html}

\bibitem{cohenthesis}
\NAME{Cohen J.}, \TITLE{{Autonomous quantum error correction with
  superconducting qubits}}, Ph.D. thesis, Ecole Normale Superieure (2017).
\newline\urlprefix\url{https://tel.archives-ouvertes.fr/tel-01545186}

\bibitem{Puri2018}
\NAME{Puri S., Grimm A., Campagne-Ibarcq P., Eickbusch A., Noh K., Roberts G.,
  Jiang L., Mirrahimi M., Devoret M.~H. \atque Girvin S.~M.}, \IN{Phys. Rev.
  X}{9}{2019}{041009}.
\newline\urlprefix\url{http://arxiv.org/abs/1807.09334
  http://dx.doi.org/10.1103/PhysRevX.9.041009
  https://link.aps.org/doi/10.1103/PhysRevX.9.041009}

\bibitem{Baragiola2019}
\NAME{Baragiola B.~Q., Pantaleoni G., Alexander R.~N., Karanjai A. \atque
  Menicucci N.~C.}, \IN{Phys. Rev. Lett.}{123}{2019}{200502}.
\newline\urlprefix\url{http://arxiv.org/abs/1903.00012
  http://dx.doi.org/10.1103/PhysRevLett.123.200502
  https://link.aps.org/doi/10.1103/PhysRevLett.123.200502}

\bibitem{Niset2008a}
\NAME{Niset J., Furasek J. \atque Cerf N.~J.}, \IN{Phys. Rev.
  Lett.}{102}{2009}{120501}.
\newline\urlprefix\url{http://arxiv.org/abs/0811.3128
  http://dx.doi.org/10.1103/PhysRevLett.102.120501
  https://link.aps.org/doi/10.1103/PhysRevLett.102.120501}

\bibitem{Giedke2002}
\NAME{Giedke G. \atque {Ignacio Cirac} J.}, \IN{Phys. Rev.
  A}{66}{2002}{032316}.
\newline\urlprefix\url{https://link.aps.org/doi/10.1103/PhysRevA.66.032316}

\bibitem{Eisert2002}
\NAME{Eisert J., Scheel S. \atque Plenio M.~B.}, \IN{Phys. Rev.
  Lett.}{89}{2002}{137903}.
\newline\urlprefix\url{http://arxiv.org/abs/quant-ph/0204052
  http://dx.doi.org/10.1103/PhysRevLett.89.137903
  https://link.aps.org/doi/10.1103/PhysRevLett.89.137903}

\bibitem{VanLoock2010}
\NAME{van Loock P.}, \IN{J. Mod. Opt.}{57}{2010}{1965}.
\newline\urlprefix\url{http://www.tandfonline.com/doi/abs/10.1080/09500340.2010.499047}

\bibitem{Hanggli2021}
\NAME{Hanggli L. \atque Konig R.}, \IN{IEEE Trans. Inf.
  Theory}{68}{2022}{1068}.
\newline\urlprefix\url{http://arxiv.org/abs/2102.05545
  http://dx.doi.org/10.1109/TIT.2021.3126881
  https://ieeexplore.ieee.org/document/9610029/}

\bibitem{Kapustin2018}
\NAME{Kapustin A. \atque Fidkowski L.}, \IN{Commun. Math.
  Phys.}{373}{2020}{763}.
\newline\urlprefix\url{http://arxiv.org/abs/1810.07756
  http://dx.doi.org/10.1007/s00220-019-03444-1
  http://link.springer.com/10.1007/s00220-019-03444-1}

\bibitem{DeMarco2021a}
\NAME{DeMarco M. \atque Wen X.-G.}, \IN{Phys. Rev. Lett.}{126}{2021}{021603}.
\newline\urlprefix\url{https://link.aps.org/doi/10.1103/PhysRevLett.126.021603}

\bibitem{DeMarco2021}
\NAME{DeMarco M. \atque Wen X.-G.}, \IN{e-print}{}{2021}{}.
\newline\urlprefix\url{http://arxiv.org/abs/2102.13057}

\bibitem{Aaronson2010}
\NAME{Aaronson S. \atque Arkhipov A.}, \TITLE{{The computational complexity of
  linear optics}}, in proc. of \TITLE{Proc. 43rd Annu. ACM Symp. Theory Comput.
  - STOC '11} (ACM Press, New York, New York, USA) 2011, p. 333.
\newline\urlprefix\url{http://arxiv.org/abs/1011.3245
  http://portal.acm.org/citation.cfm?doid=1993636.1993682}

\bibitem{Wang2019}
\NAME{Wang C.~S., Curtis J.~C., Lester B.~J., Zhang Y., Gao Y.~Y., Freeze J.,
  Batista V.~S., Vaccaro P.~H., Chuang I.~L., Frunzio L., Jiang L., Girvin
  S.~M. \atque Schoelkopf R.~J.}, \IN{Phys. Rev. X}{10}{2020}{021060}.
\newline\urlprefix\url{http://arxiv.org/abs/1908.03598
  http://dx.doi.org/10.1103/PhysRevX.10.021060
  https://link.aps.org/doi/10.1103/PhysRevX.10.021060}

\bibitem{Oh2022}
\NAME{Oh C., Lim Y., Wong Y., Fefferman B. \atque Jiang L.},
  \IN{e-print}{}{2022}{}.
\newline\urlprefix\url{http://arxiv.org/abs/2202.01861}

\bibitem{Arenz2018}
\NAME{Arenz C., Bondar D.~I., Burgarth D., Cormick C. \atque Rabitz H.},
  \IN{Quantum}{4}{2020}{271}.
\newline\urlprefix\url{http://arxiv.org/abs/1806.00444
  http://dx.doi.org/10.22331/q-2020-05-25-271
  https://quantum-journal.org/papers/q-2020-05-25-271/}

\bibitem{SivakVolodymyr}
\NAME{Sivak V., Eickbusch A., Royer B., Tsioutsios I., Schoelkopf R.~J. \atque
  Devoret M.~H.}, \TITLE{{Extending the logical lifetime of a stabilized
  Gottesman-Kitaev-Preskill qubit with reinforcement learning}} (2021).
\newline\urlprefix\url{https://meetings.aps.org/Meeting/MAR22/Session/N37.2}

\bibitem{Ryan-Anderson2022}
\NAME{Ryan-Anderson C., Brown N.~C., Allman M.~S., Arkin B., Asa-Attuah G.,
  Baldwin C., Berg J., Bohnet J.~G., Braxton S., Burdick N., Campora J.~P.,
  Chernoguzov A., Esposito J., Evans B., Francois D., Gaebler J.~P., Gatterman
  T.~M., Gerber J., Gilmore K., Gresh D., Hall A., Hankin A., Hostetter J.,
  Lucchetti D., Mayer K., Myers J., Neyenhuis B., Santiago J., Sedlacek J.,
  Skripka T., Slattery A., Stutz R.~P., Tait J., Tobey R., Vittorini G., Walker
  J. \atque Hayes D.}, \IN{e-print}{}{2022}{}.
\newline\urlprefix\url{http://arxiv.org/abs/2208.01863}

\bibitem{Chen2021a}
\NAME{Chen Z., Satzinger K.~J., Atalaya J., Korotkov A.~N., Dunsworth A., Sank
  D., Quintana C., McEwen M., Barends R., Klimov P.~V., Hong S., Jones C.,
  Petukhov A., Kafri D., Demura S., Burkett B., Gidney C., Fowler A.~G., Paler
  A., Putterman H., Aleiner I., Arute F., Arya K., Babbush R., Bardin J.~C.,
  Bengtsson A., Bourassa A., Broughton M., Buckley B.~B., Buell D.~A., Bushnell
  N., Chiaro B., Collins R., Courtney W., Derk A.~R., Eppens D., Erickson C.,
  Farhi E., Foxen B., Giustina M., Greene A., Gross J.~A., Harrigan M.~P.,
  Harrington S.~D., Hilton J., Ho A., Huang T., Huggins W.~J., Ioffe L.~B.,
  Isakov S.~V., Jeffrey E., Jiang Z., Kechedzhi K., Kim S., Kitaev A.,
  Kostritsa F., Landhuis D., Laptev P., Lucero E., Martin O., McClean J.~R.,
  McCourt T., Mi X., Miao K.~C., Mohseni M., Montazeri S., Mruczkiewicz W.,
  Mutus J., Naaman O., Neeley M., Neill C., Newman M., Niu M.~Y., O'Brien
  T.~E., Opremcak A., Ostby E., Pat{\'{o}} B., Redd N., Roushan P., Rubin
  N.~C., Shvarts V., Strain D., Szalay M., Trevithick M.~D., Villalonga B.,
  White T., Yao Z.~J., Yeh P., Yoo J., Zalcman A., Neven H., Boixo S.,
  Smelyanskiy V., Chen Y., Megrant A. \atque Kelly J.},
  \IN{Nature}{595}{2021}{383}.
\newline\urlprefix\url{http://arxiv.org/abs/2102.06132
  http://dx.doi.org/10.1038/s41586-021-03588-y
  http://www.nature.com/articles/s41586-021-03588-y}

\bibitem{Acharya2022}
\NAME{Acharya R., Aleiner I., Allen R., Andersen T.~I., Ansmann M., Arute F.,
  Arya K., Asfaw A., Atalaya J., Babbush R., Bacon D., Bardin J.~C., Basso J.,
  Bengtsson A., Boixo S., Bortoli G., Bourassa A., Bovaird J., Brill L.,
  Broughton M., Buckley B.~B., Buell D.~A., Burger T., Burkett B., Bushnell N.,
  Chen Y., Chen Z., Chiaro B., Cogan J., Collins R., Conner P., Courtney W.,
  Crook A.~L., Curtin B., Debroy D.~M., Barba A. D.~T., Demura S., Dunsworth
  A., Eppens D., Erickson C., Faoro L., Farhi E., Fatemi R., Burgos L.~F.,
  Forati E., Fowler A.~G., Foxen B., Giang W., Gidney C., Gilboa D., Giustina
  M., Dau A.~G., Gross J.~A., Habegger S., Hamilton M.~C., Harrigan M.~P.,
  Harrington S.~D., Higgott O., Hilton J., Hoffmann M., Hong S., Huang T., Huff
  A., Huggins W.~J., Ioffe L.~B., Isakov S.~V., Iveland J., Jeffrey E., Jiang
  Z., Jones C., Juhas P., Kafri D., Kechedzhi K., Kelly J., Khattar T., Khezri
  M., Kieferov{\'{a}} M., Kim S., Kitaev A., Klimov P.~V., Klots A.~R.,
  Korotkov A.~N., Kostritsa F., Kreikebaum J.~M., Landhuis D., Laptev P., Lau
  K.-M., Laws L., Lee J., Lee K., Lester B.~J., Lill A., Liu W., Locharla A.,
  Lucero E., Malone F.~D., Marshall J., Martin O., McClean J.~R., Mccourt T.,
  McEwen M., Megrant A., Costa B.~M., Mi X., Miao K.~C., Mohseni M., Montazeri
  S., Morvan A., Mount E., Mruczkiewicz W., Naaman O., Neeley M., Neill C.,
  Nersisyan A., Neven H., Newman M., Ng J.~H., Nguyen A., Nguyen M., Niu M.~Y.,
  O'Brien T.~E., Opremcak A., Platt J., Petukhov A., Potter R., Pryadko L.~P.,
  Quintana C., Roushan P., Rubin N.~C., Saei N., Sank D., Sankaragomathi K.,
  Satzinger K.~J., Schurkus H.~F., Schuster C., Shearn M.~J., Shorter A.,
  Shvarts V., Skruzny J., Smelyanskiy V., Smith W.~C., Sterling G., Strain D.,
  Szalay M., Torres A., Vidal G., Villalonga B., Heidweiller C.~V., White T.,
  Xing C., Yao Z.~J., Yeh P., Yoo J., Young G., Zalcman A., Zhang Y. \atque Zhu
  N.}, \IN{e-print}{}{2022}{}.
\newline\urlprefix\url{http://arxiv.org/abs/2207.06431}

\end{thebibliography}

\end{document}